%% file: exposure.tex
\documentclass[11pt, a4paper]{article}
\usepackage[latin1]{inputenc}
\usepackage{fancyhdr}
\usepackage{graphics,epsfig,color}
\usepackage{wrapfig,rotating}
\usepackage{amssymb,amsmath,array}
\usepackage{epstopdf}
\usepackage{lscape}
\usepackage{url}
\usepackage{afterpage}
\usepackage{units}
\usepackage{endnotes}
\usepackage{footmisc}
\newcommand{\eplus}{\mathrm{e}^+}
\newcommand{\eminus}{\mathrm{e}^-}
\newcommand{\epem}{\eplus\eminus}

\newcommand{\ttbar}{\mathrm{t}\overline{\mathrm{t}}}

\newcommand{\roots}{\sqrt{s}}

\newcommand{\balpha}{\boldsymbol{\alpha}}
\DefineFNsymbols*{asterisks}{*{$\dagger$}{$\ddagger$}{$\mathchar "278$}{$\mathchar "27B$}{$\|$}{**}{$\dagger\dagger$}{$\ddagger\ddagger$}{$\mathchar"27C$}}
\setfnsymbol{asterisks}

\usepackage[]{lineno}
\linenumbers
\begin{document}
\title{
Effects of high-energy particle showers on the embedded front-end electronics of an electromagnetic calorimeter for a future lepton collider } 
\author{\centering \LARGE\bf The CALICE Collaboration}
\date{}

\maketitle
\thispagestyle{fancy}

\begin{abstract}
{\em A}pplication {\em S}pecific {\em I}ntegrated {\em C}ircuits, ASICs, similar to those envisaged for the readout electronics of the central calorimeters of detectors for a future lepton collider have been exposed to high-energy electromagnetic showers. A salient feature of these calorimeters is that the readout electronics will be embedded into the calorimeter layers. In this article it is shown that interactions of shower particles in the volume of the readout electronics do not alter the noise pattern of the ASICs.
No signal at or above the MIP level has been observed during the exposure. The upper limit at the 95\% confidence 
level on the frequency of fake signals is smaller than $1\cdot10^{-5}$ for a noise threshold of about 60\% of a MIP. For ASICs with similar design to those which were tested, it can thus be largely excluded that the embedding of the electronics into the calorimeter layers compromises the performance of the calorimeters.
\end{abstract}


\newpage
\input{authors_inc}

\pagenumbering{arabic}
\setcounter{page}{1}
\section{Introduction}
The central calorimeters of the detectors to be operated at a future lepton collider will have the readout  electronics embedded into the active layers of the calorimeter~\cite{ild09, sid-loi, fourth}. The energy of electromagnetic showers produced  in the final states ranges between a few MeV up to several hundreds of GeV. A natural question arising from this design is whether the cascade particles of the  high-energy showers which penetrate through the electronics do create radiation induced effects in these circuits.  These effects would compromise the precision measurements envisaged at the lepton collider. Possible radiation effects include Transient Effects and Single Event Upsets~\cite{hre} which may create pulses which would be recorded as {\em fake signals} or, even worse, could cause damage to the readout electronics. 

The CALICE collaboration is designing, building and operating large scale prototypes for the calorimeters at a future lepton collider~\cite{imad}. Large statistics data samples have been recorded in test beam campaigns in order to understand the behaviour of highly granular calorimeters. This article describes the measurements conducted in a special set of runs in which an ordinary calorimeter layer of a prototype 
for a silicon tungsten electromagnetic calorimeter, called {\em SiW Ecal} hereafter, has been replaced by a {\em special PCB} allowing for the exposure of the readout electronics to particle showers. The data analysed here were recorded during the 2007 test beam campaign at CERN in the H6 test beam area

\section{Experimental set-up and data taking}

\begin{figure}[h!]
\begin{minipage}[l]{0.45\columnwidth}
\centerline{\includegraphics[width=1.09\columnwidth]{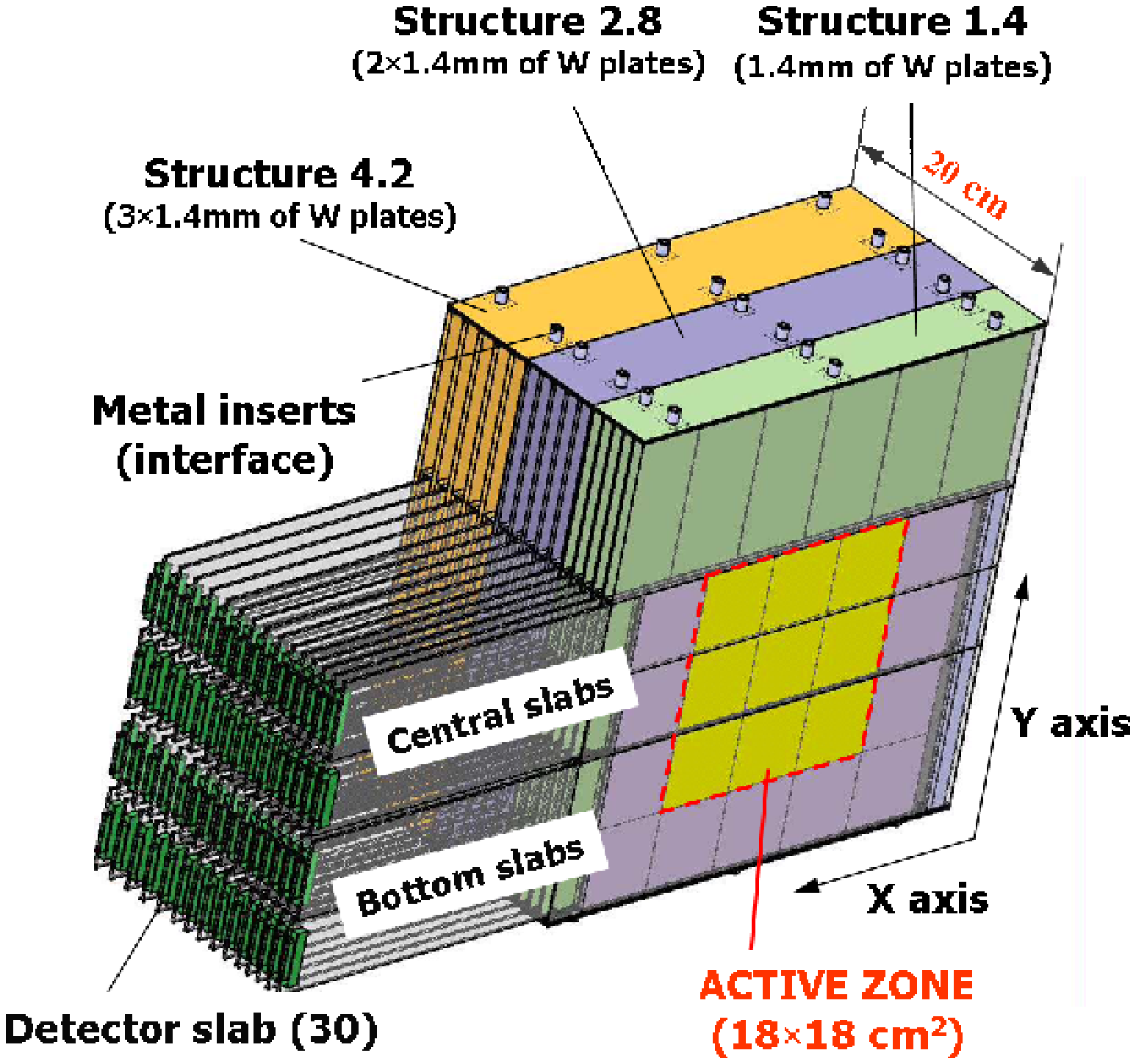}}
\caption{\sl Schematic 3D view of the physics prototype.}
\label{fig:3Dproto}
\end{minipage}
\hfill
\begin{minipage}[l]{0.45\columnwidth}
\centerline{\includegraphics[width=1.09\columnwidth]{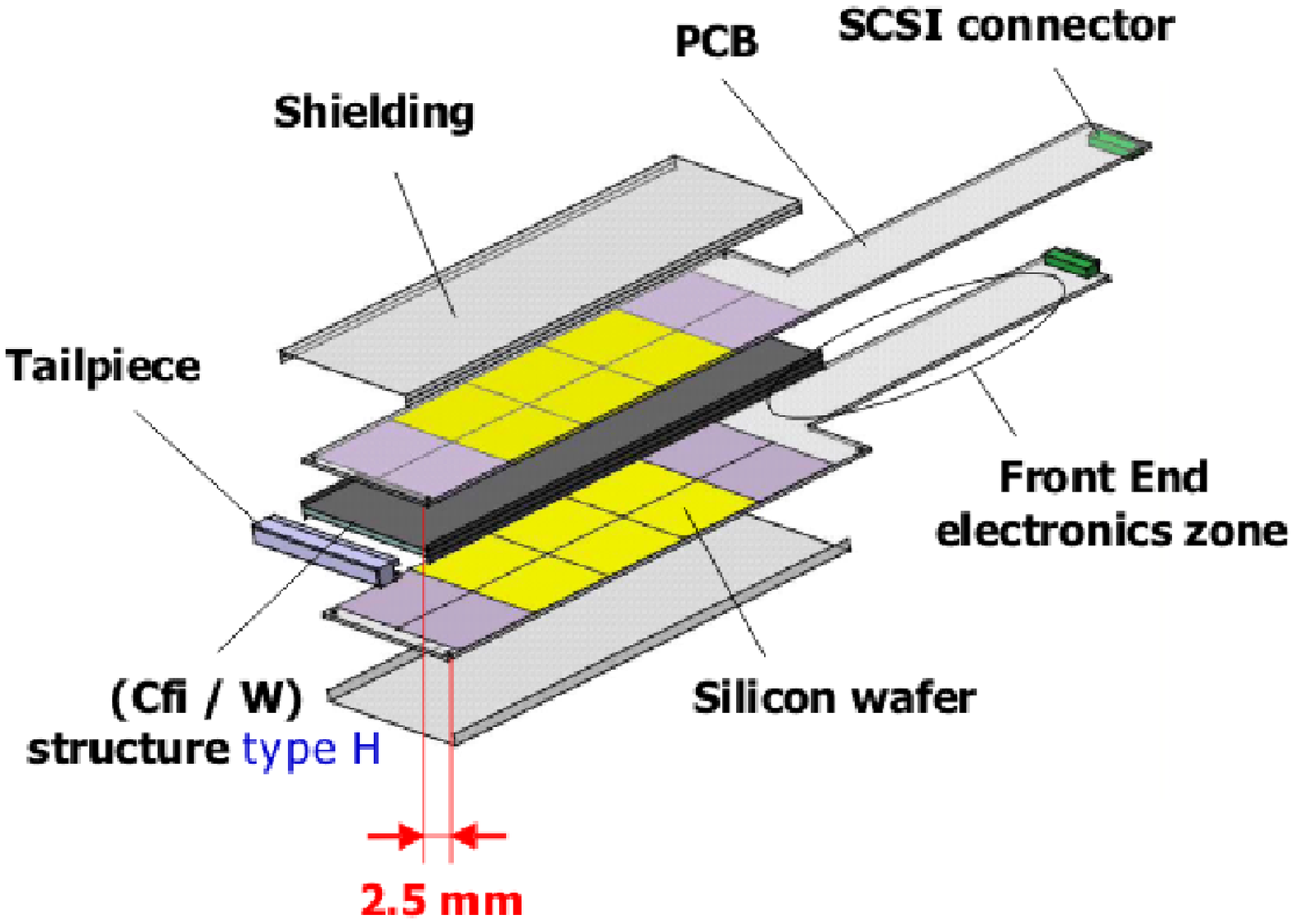}}
\caption{\sl Schematic diagram showing the components of a detector slab.}
\label{fig:slab}
\end{minipage}
\end{figure}

Figure~\ref{fig:3Dproto} shows a perspective view on the {\it physics prototype} of the SiW Ecal. A comprehensive description of the physics prototype is given elsewhere~\cite{calice1}. Here only those 
features relevant for the present analysis will be outlined.

\begin{figure}[htpb]
\centering
\includegraphics[width=0.6\textwidth]{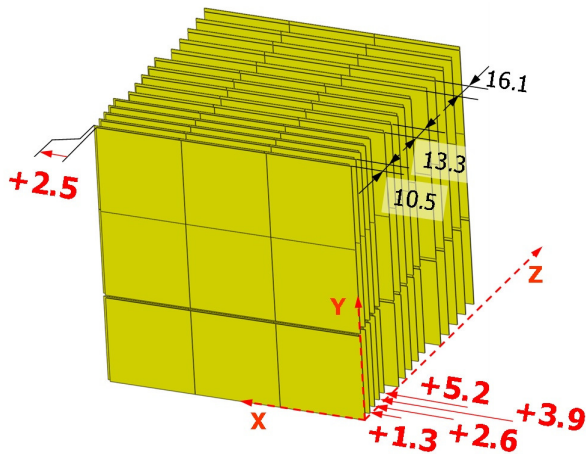}%
\caption{\sl Illustration of layer offsets (2.5\,mm) and slabs offsets (1.3\,mm) of the physics prototype of the SiW Ecal as discussed in the text. Shown is also the total extension of the three modules in $z$-direction. All dimensions are in mm.}
\label{fig:offandrot}
\end{figure}

The physics prototype consists of 9720 1x1\,${\rm cm^2}$ wide calorimeter cells subdivided into 30 layers. The active zone covers ${\rm 18 \times 18\,cm^2}$ in width and approximately 20\,cm in depth. 
The layers are composed alternately  by W absorber plates and a matrix of PIN diode sensors on a silicon wafer substrate. At normal incidence, the prototype has a total depth of 24\,$X_0$ achieved using 10 layers of 0.4\,$X_0$ 
tungsten absorber plates, followed by 10 layers of 0.8\,$X_0$, and another 10 layers of 1.2\,$X_0$ thick plates. 
Each layer is subdivided into a central part featuring a $3\times2$ array of silicon wafers and a bottom part consisting of a $3\times1$ 
array of silicon wafers. Note that in the running period relevant for this analysis the bottom part of the first six layers was missing. 

The silicon wafers are mounted onto both sides of an H-shaped tungsten plate as shown in Figure~\ref{fig:slab}. Such an entity is called a {\em slab}. In order to avoid an alignment of wafer boundaries, the layers within a slab are shifted 
by 2.5\,mm in the positive $x$-direction with respect to each other. In the same way, two successive slabs are shifted by 1.3\,mm with respect to each other. The layer offsets are illustrated in Figure~\ref{fig:offandrot}.

\begin{figure}[h!]
\centerline{\includegraphics[width=0.5\columnwidth]{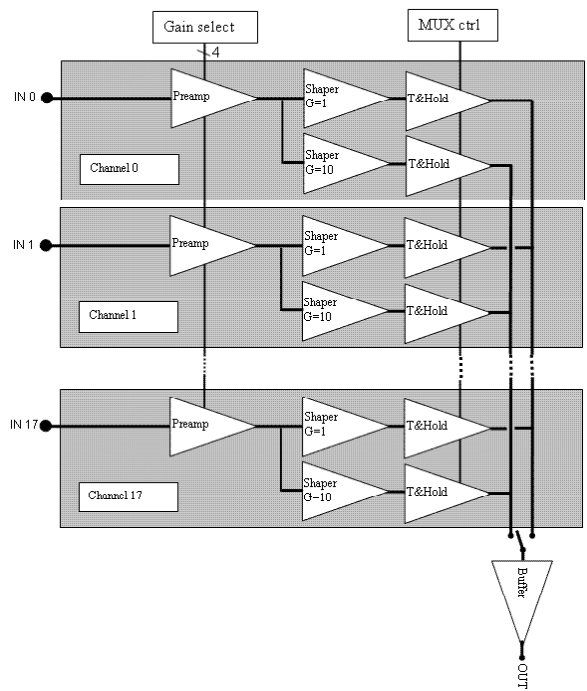}}
\caption{\sl General block schematic of FLC\_PHY3.}
\label{fig:block}
\end{figure}

As indicated in Figures~\ref{fig:3Dproto} and~\ref{fig:slab}, the readout electronics are located outside the absorber structure and hence not exposed to high-energy electromagnetic showers. The main device of the read out electronics is an 18 channel charge sensitive ASIC, called FLC\_PHY3 which is realised in 0.8\,\u{\um} AMS BiCMOS technology. One $6\times6$ sensor matrix is thus read out by two ASICs. This provides redundancy by de-correlating ASIC and sensor response. As shown in Figure~\ref{fig:block}, the signal path starts with a variable gain charge preamplifier, followed by two shaping filters for gains 1 and 10 with a shaping time of 180\,ns for both gains. The shaped signal is then propagated to a sample and hold device realised by a 2\,pF capacitance. After that the measured voltage, which is proportional 
to the charge at the input of the pre-amplifier, is written into a buffer designed to store the 18 signals as processed by the signal chain. The 18 signals are read one-by-one by the off-detector electronics. One channel covers a dynamic range equivalent to the energy deposition by about 600\,{\em M}inimum {\em I}onising {\em P}articles, MIPs, which has been considered to be sufficient for a beam test using primary electrons of an energy of up to 50\,GeV.

For the present tests, FLC\_PHY3 ASICs were exposed to electromagnetic showers. One ASIC has a
surface of about $1.6\times2.3\,\rm{mm^2}$. It is TQFP64 packaged such that the whole ensemble has outer dimensions of about $1\times1\,\rm{cm^2}$~\cite{dlt09}. The shower particles may create charges
and thus fake signals in the PMOS at the entrance of the pre-amplifiers of the 18 channels. The sensitive area of one channel 
is about $3000\u{\um^2}$ while the total surface of a channel is about $110000\,\rm{\mu m^2}$. Signals created in the circuitry after the pre-amplifier which would appear immediately at the output of the ASIC cannot be recorded due to the sampling latency of 180\,ns of the CALICE data acquisition system. Radiation effects could therefore only become apparent in case of a failure of the circuitry. Such a failure has not been observed during the tests presented in this article.

\begin{figure}[htpb]
\centering
\includegraphics[width=0.48\textwidth]{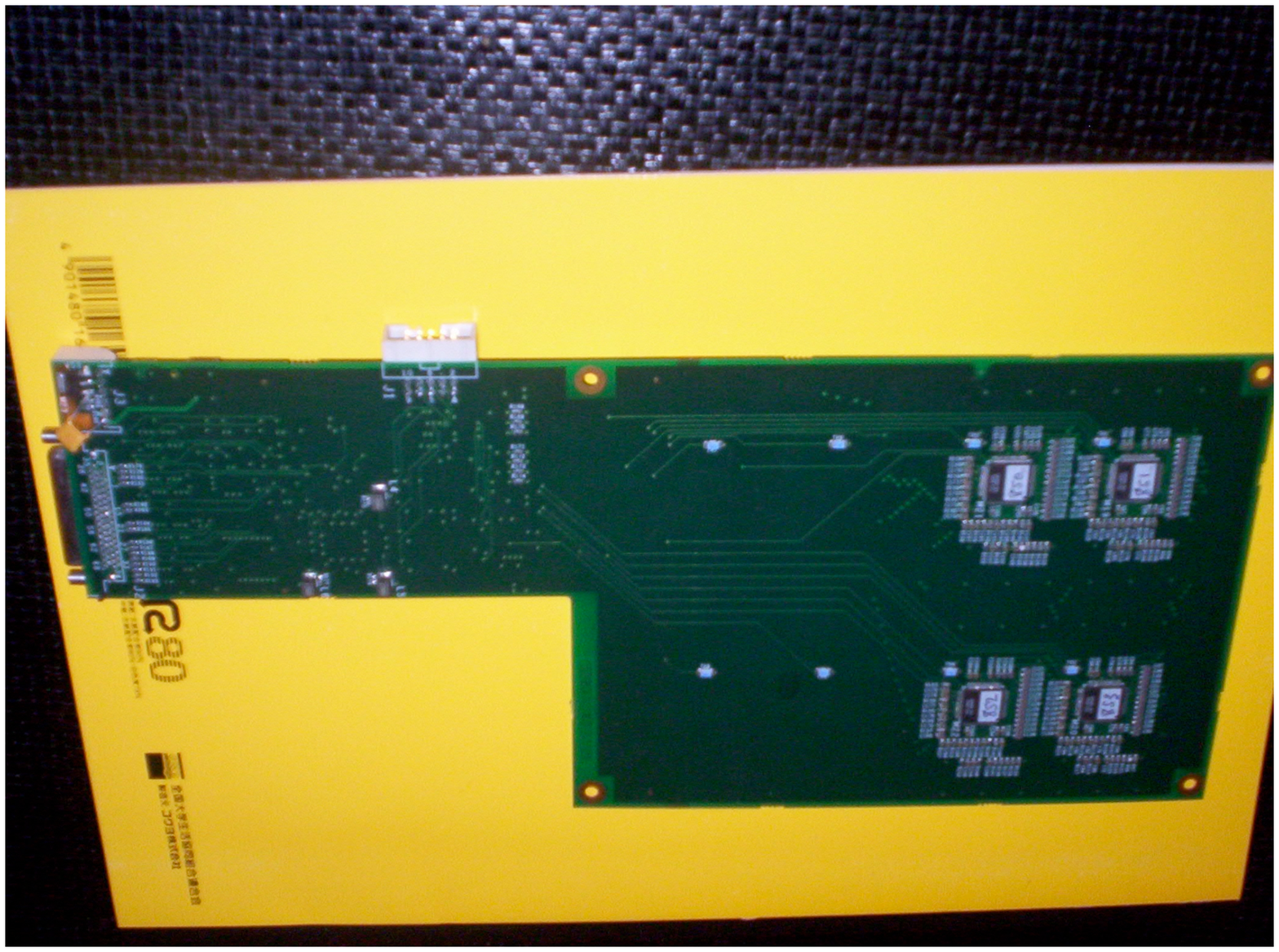}%
\includegraphics[width=0.475\textwidth]{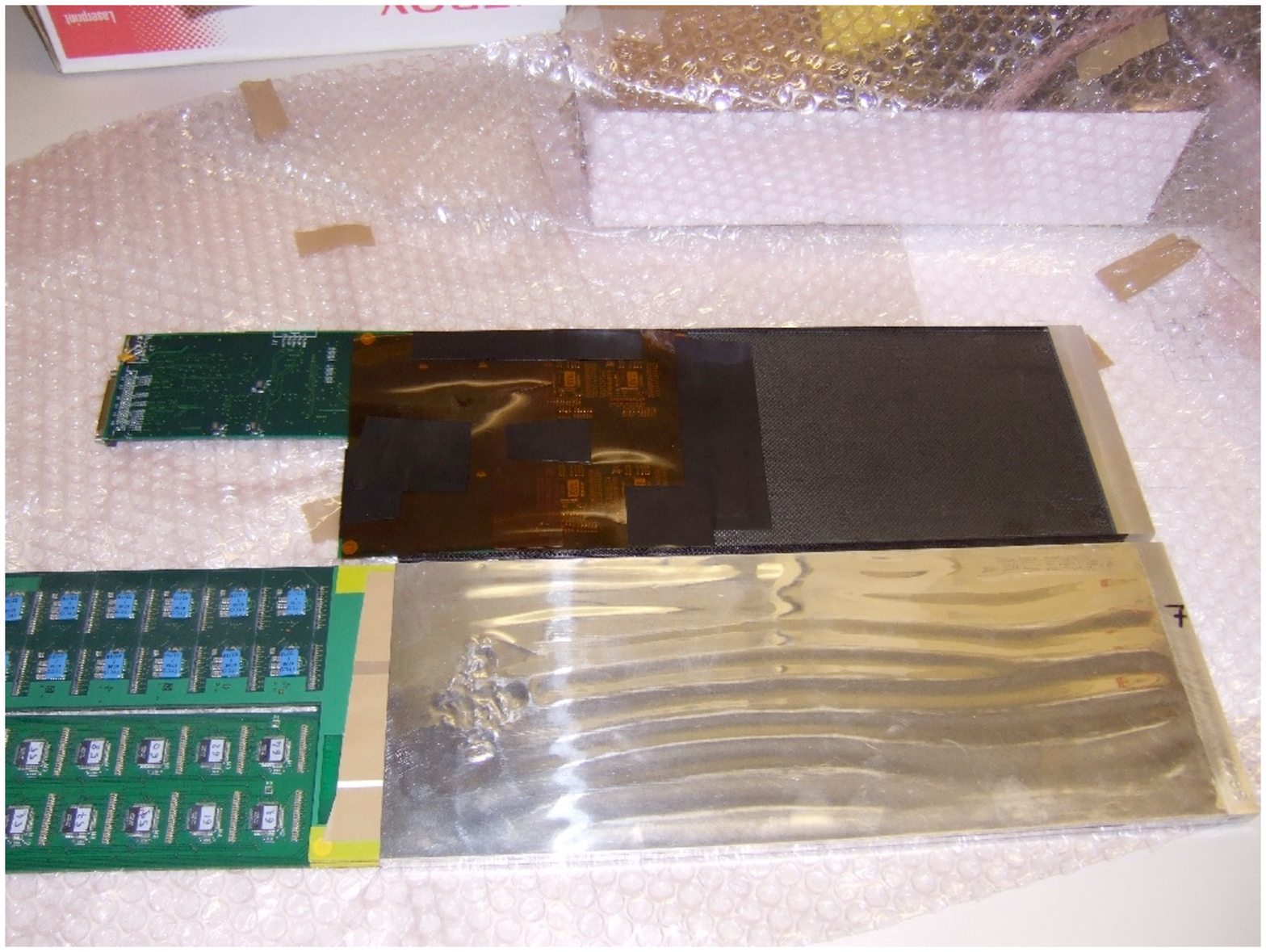}%
\caption{\sl Left: Picture of the special PCB with the four ASICs used in the test. Right: The special PCB mounted on tungsten  absorber and comparison with a regular slab.}
\label{fig:specboard}
\end{figure}

\begin{figure}[h]
\begin{center}
\includegraphics[width=0.6\textwidth, angle=0]{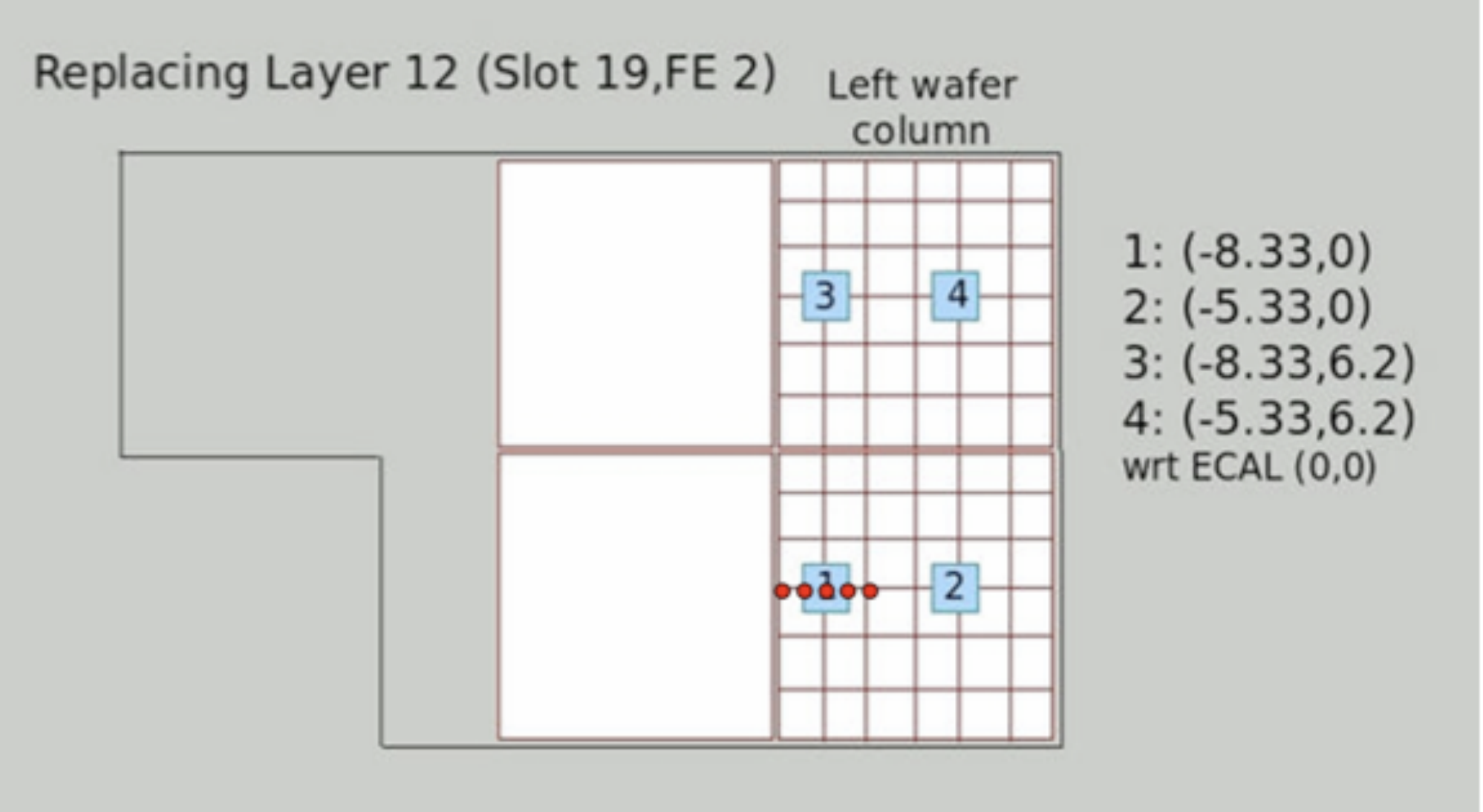}
\caption{{\sl Schematic view of the special PCB. The red points indicate the nominal five impact points for a scan over ASIC 1. Identical scans have been performed for the other three ASICs.}}
\label{fig:boardscan}
\end{center}
\end{figure}

The special PCB is equipped with four ASICs in the nominal sensitive plane of the detector. It has been mounted directly on a spare H-board as shown in Figure~\ref{fig:specboard}.
The special PCB has been placed within the physics prototype at the layer corresponding to the expected position of the shower maximum. In this configuration data with electrons with an energy of 70\,GeV and 90\,GeV have been recorded. The lateral spread of the electron beam at these energies is about 1\,cm in diameter~\cite{calice1}. The beam has been positioned at five different points along the $x$-direction at the centre in $y$ of each of the four ASICs as indicated for the 'ASIC 1' in Figure~\ref{fig:boardscan}. 
Beam events triggered with scintillation counters, called {\em signal events} hereafter, are interleaved during the data taking with {\em pedestal events} triggered by an oscillator integrated into the CALICE data acquisition system. For further details of the experimental set-up please consult~\cite{calice1, calice2, calice3}. Table~\ref{tab:protocol} gives the number of recorded signal and pedestal events at each measurement point. The table also introduces the nomenclature used hereafter to identify the various measurement points. 


\begin{table}[htdp]
\begin{center}
\begin{footnotesize}
  \begin{tabular}{@{} |cccccc| @{}}
    \hline
    Scan & Beam centred on& Measurement point & Position: x,y\,[cm]& Signal events & Pedestal events \\ 
    \hline
      & & 1&-9.33, 0 & 78293 & 14624\\ 
      & & 2& -8.83, 0& 189966 &  37173\\ 
     1& ASIC 1 & 3 & -8.33, 0 & 209312 & 38361 \\ 
      & & 4& -7.83, 0 & 65249 & 3602 \\ 
      & & 5&-7.33, 0 & 85543 & 4306 \\ 
    \hline
     & & 1 & -6.33, 0 & 85188 & 4678 \\ 
     & & 2 & -5.83, 0 & 129778 & 6146\\ 
    2& ASIC 2 & 3 & -5.33, 0 & 213369 & 13719 \\ 
     & & 4 & -4.83, 0 & 217111 & 11053 \\ 
     & & 5 & -4.33, 0 & 89435 & 4254 \\ 
     \hline
     & & 1 & -9.33, 6.2 & 90395 & 4347 \\ 
     & & 2 & -8.83, 6.2 & 228138 & 10296 \\ 
    3 & ASIC 3 & 3 & -8.33, 6.2 & 216877 & 9831 \\ 
     & & 4 & -7.83, 6.2 & 218519 & 9462 \\ 
     & & 5 & -7.33, 6.2 & 86989 & 3909 \\ 
    \hline
     & & 1 & -6.33, 6.2 & 66655 & 4223 \\ 
     & & 2 & -5.83, 6.2 & 214418 & 13666 \\ 
    4 & ASIC 4  & 3 & -5.33, 6.2 & 314275 & 15264 \\ 
     & & 4 & -4.83, 6.2 & 217415 & 11698 \\ 
     & & 5 & -4.33, 6.2 & 85884 & 4949 \\ 
    \hline
  \end{tabular}
\end{footnotesize}
\end{center}
\caption{\sl Protocol of the exposure test containing the identifiers of the measurement points, their position and the number of signal and
pedestal events recorded at each position.}
\label{tab:protocol}
\end{table}%

\section{Initial steps of data analysis}

\begin{figure}
\begin{center}
\includegraphics[width=0.95\textwidth]{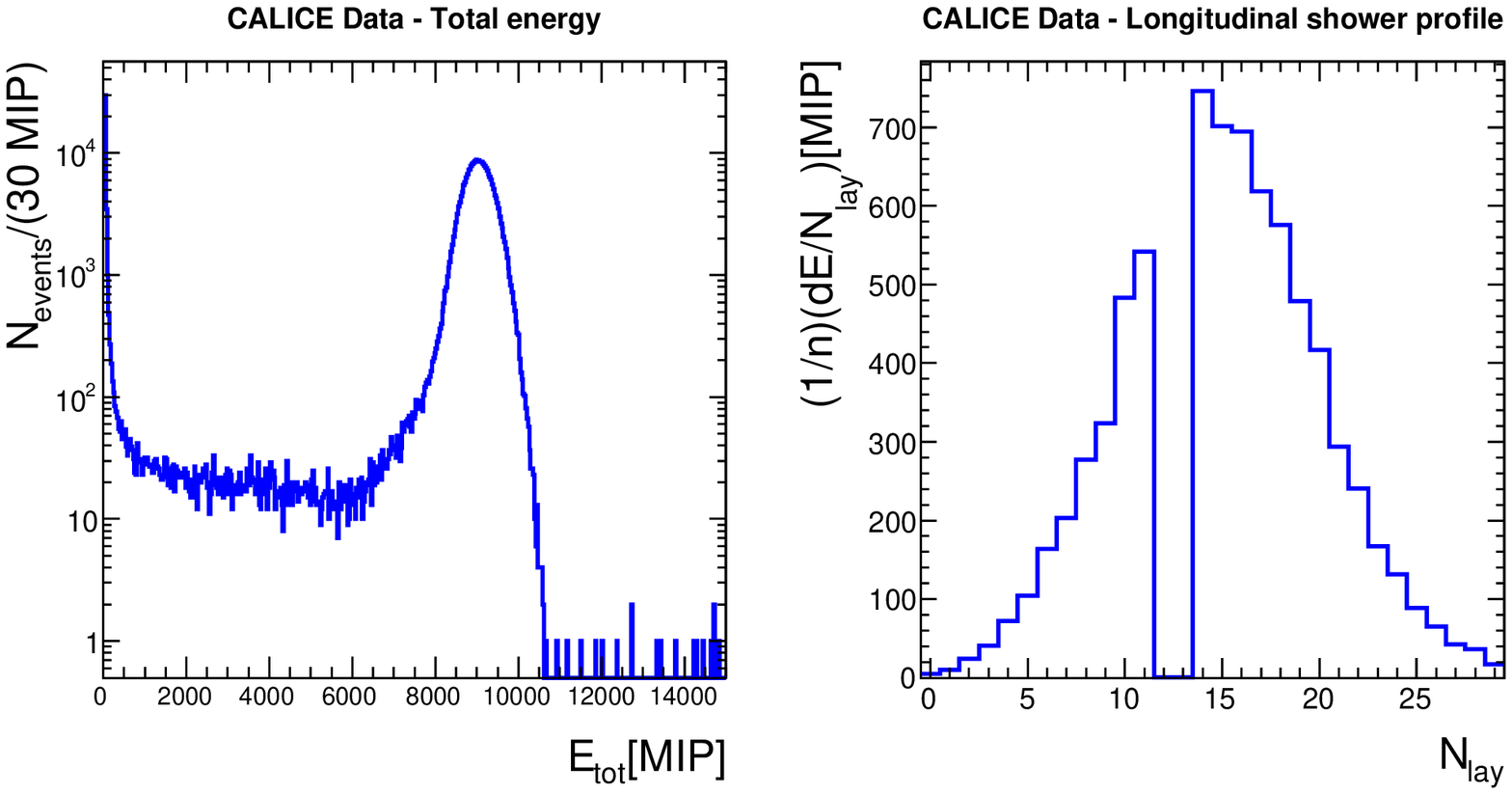}
\caption{\sl Total energy deposition and longitudinal shower profile for a run with electrons of 90\,GeV during the test with the special PCB.
The gap visible in the longitudinal shower profile is due to the replacement of a regular slab by the H-board carrying the special PCB.}
\label{fig:eneglong}
\end{center}
\end{figure}

The data are verified for a proper alignment of the ASICs relative to the beam in lateral direction and to the shower maximum in longitudinal direction. 
Figure~\ref{fig:eneglong} shows the  spectra of a run with electrons of 90\,GeV. Here, the recorded data were reconstructed with the same reconstruction chain as applied to the regular data taking~\cite{calice1, calice2}. The energy deposition in the detector 
is given in terms of MIPs and 1\,MIP corresponds to about 
45\,{\em ADC counts}~\cite{calice1} as recorded by the CALICE data acquisition system. The reconstruction 
chain introduces a zero suppression at 0.6\,MIP corresponding to approximately 4.5 times the mean noise level of 6 ADC counts. 
In addition, a correction for pedestal instabilities caused by insufficient isolation of the power supply lines of the PCBs 
is applied~\cite{calice1}. After this correction the residual pedestal instability is about 0.2\% of a MIP (or 0.1\,ADC counts).

\begin{figure}
\begin{center}
\includegraphics[width=0.95\textwidth]{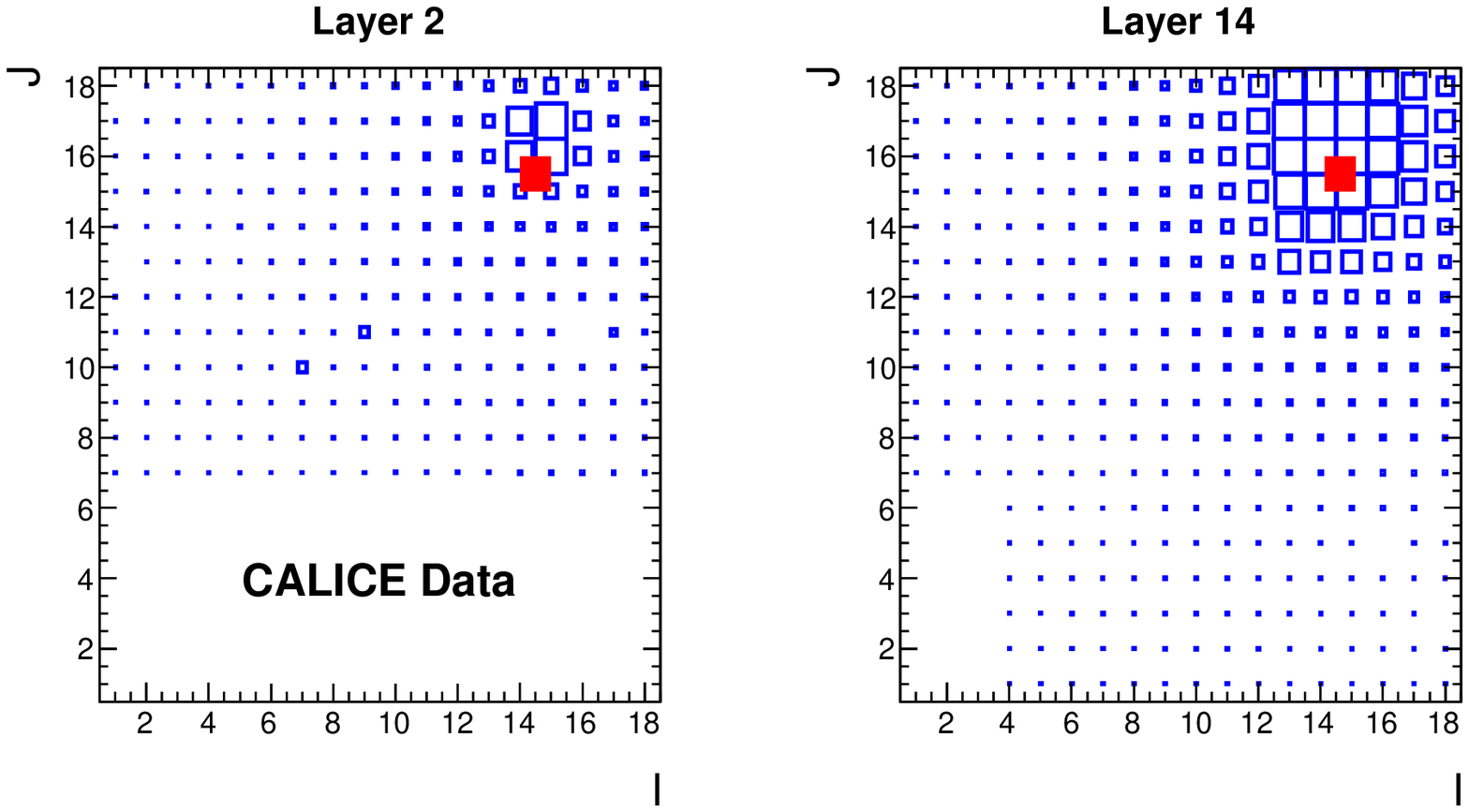}
\caption{\sl Hit maps shown as a function of the cell indices $I,\,J$ in x and y-directions in layer 2 and 14.
The area of the boxes represents the number of recorded hits in the individual cells.
The position of ASIC 4, indicated by the full square, is projected into the planes of these layers.}
\label{fig:align}
\end{center}
\end{figure}

The total energy spectrum exhibits a clear maximum well separated from residual noise and MIP events. The gap visible in the 
longitudinal shower profile indicates that the special layer has been placed close to the shower maximum.  
The lateral position of the special PCB, installed at the position of Layer 12, is 
identical to that of the Layer 2 of the prototype. As an example, Figure~\ref{fig:align} shows the hit maps of layers 2 and 14 for a run in which the beam was incident on ASIC 4. The Layer 14 is the first regular layer behind the special PCB. Overlaid to the hit maps is the projected position of ASIC 4. The gaps in the lower parts of the hit maps can be explained by non-instrumented parts of the detector.

It is clearly  visible that the beam hits the detector close to the ASIC  position and that the lateral shower development leads to a 
good coverage and thus good exposure to shower particles of the ASIC.
 
\begin{figure}
\begin{center}
\includegraphics[width=0.7\textwidth]{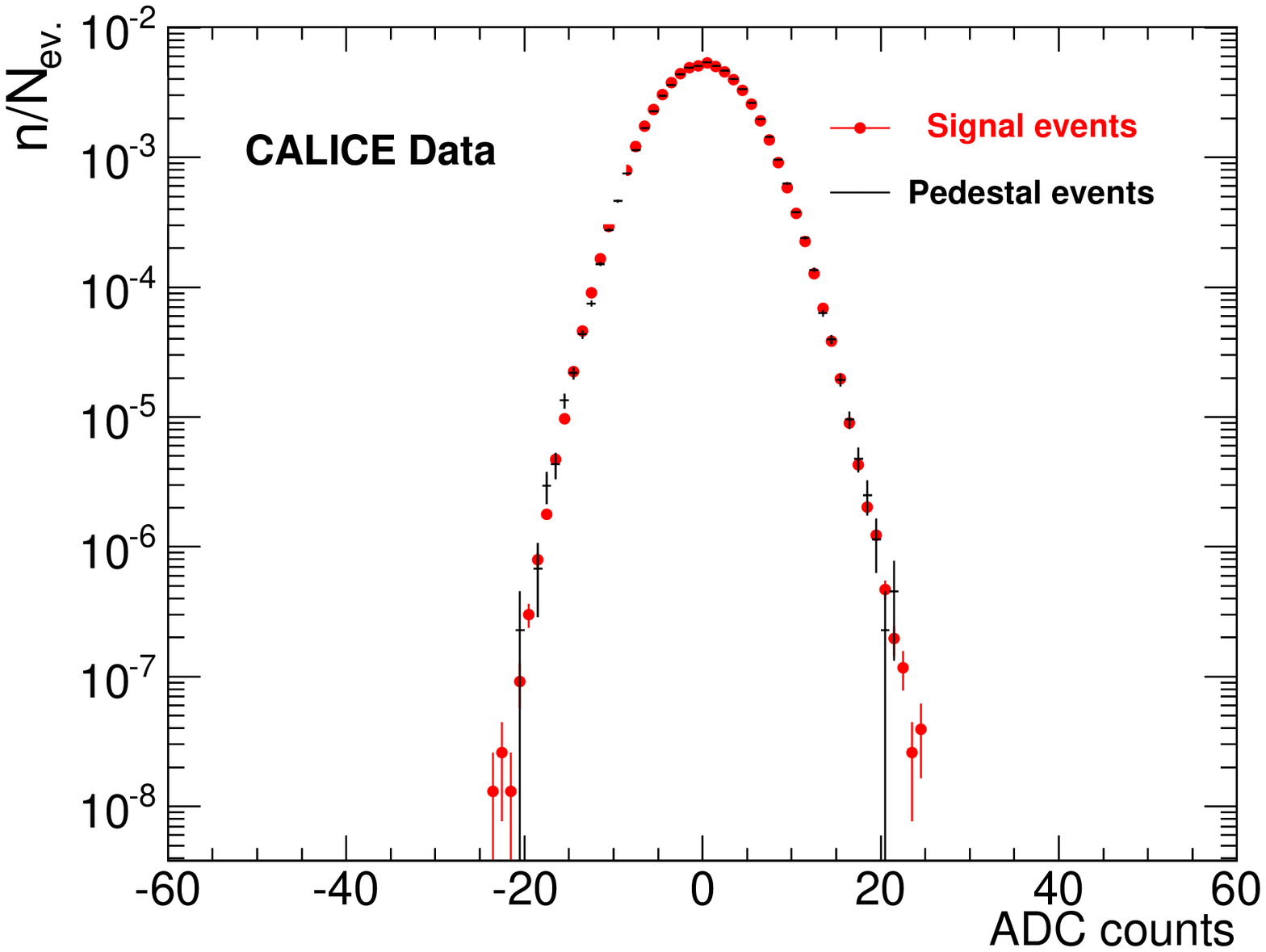}
\caption{\sl Noise spectrum for signal events and pedestal events. The figure displays the result for the run in which the beam was centred on ASIC 4. The distributions are normalised to unity.}
\label{fig:cscanhist4}
\end{center}
\end{figure}

\begin{figure}[ht]
\begin{center}
\includegraphics[width=0.9\textwidth]{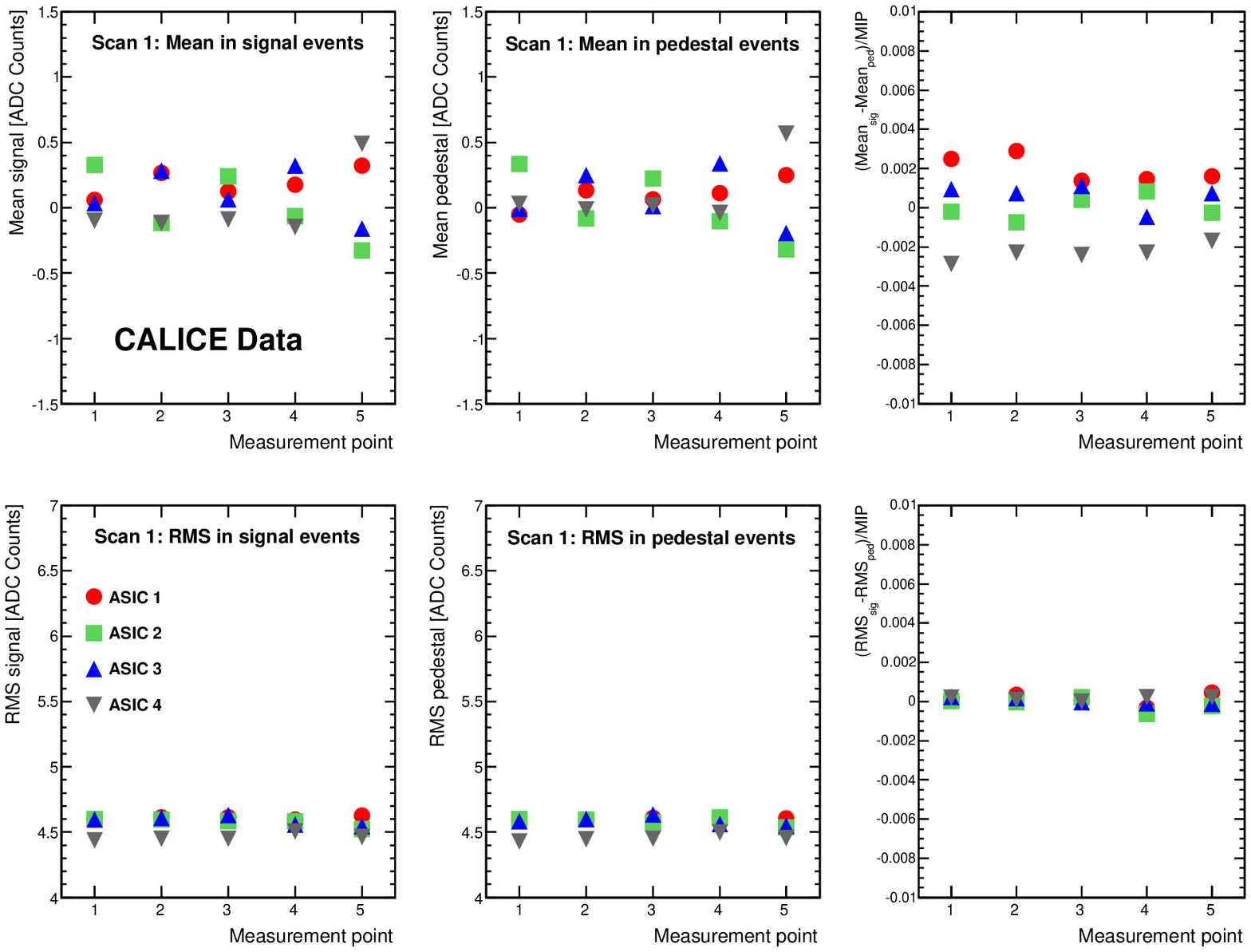}
\caption{\sl Mean and RMS for signal events and pedestal events. The very right part shows the corresponding differences normalised to the value of a MIP assumed to be 45 ADC counts. The figure displays the result for Scan 1 in which ASIC 1 is scanned. As a cross-check the results for all ASICs are shown.}
\label{fig:cscan1}
\end{center}
\end{figure}

In a next step the regular zero suppression was switched off in the CALICE reconstruction program in order to be sensitive 
to the behaviour of the ASICs in the small signal range. For technical reasons the first channel  
of each ASIC on the special PCB is discarded, leaving 17 signals per ASIC per event. As motivated by the energy spectrum shown in Figure~\ref{fig:eneglong}, the signal events are further selected 
by requiring an energy deposition of more than 2000 MIPs in order to be unbiased by MIP-like events.
This cut reduces the available statistics quoted in Table~\ref{tab:protocol} by approximately 15\%.
Still, no difference between the noise spectra obtained for signal and pedestal events is expected. As an example, in Figure~\ref{fig:cscanhist4} 
the noise spectra of signal and pedestal events are compared the Measurement point 3 of Scan 4. 
Indeed, no difference between the two data types can be observed. After this initial qualitative comparison, the mean and the root mean square (RMS) for each ASIC at 
each measurement point  are extracted for signal and pedestal events. The results are displayed in Figure~\ref{fig:cscan1}, using the scan over ASIC 1 as an example.

The corresponding figures for the other scans are given in the appendix.
From these figures the following conclusions can be drawn:
\begin{itemize}
\item The mean and the RMS remain the same throughout all the runs. In particular no dependence on the scan position can be observed.
\item The mean and RMS for signal and pedestal events are always nearly identical. Residual differences are smaller than 0.4\% of a MIP.  
\end{itemize}

\section{Detailed noise analysis}

The high-energy showers penetrating the electronics may disturb the noise characteristics of the exposed ASICs. This perturbation may be revealed in changes of the coherent and incoherent noise levels of the ASICs. A very robust and widely used technique to analyse noise patterns in data is given by a {\em Principal Component Analysis (PCA)}. The analysis performed in this paper follows that presented in~\cite{ATLAS-Note}. As was pointed out there and confirmed in present study, the PCA leads to more reliable results than a more simplistic approach based on direct and alternating sums. 

\subsection{Principal Component Analysis - PCA}

The PCA can be subdivided into five steps which are introduced now. Each step will be illustrated by the results obtained for nominal central impact on the ASICs.
\begin{enumerate}
\item The vector of noise hits $\mathbf{b}$ for a given ASIC can be decomposed into 
\begin{equation}
\mathbf{b}=\mathbf{u}+c\boldsymbol{\alpha},
\label{eq:noisevec}
\end{equation}
where $\mathbf{u}$ represents the contribution of the incoherent noise. 
The vector $\balpha$ characterises the correlation among the ASICs. More specifically, its components quantify the relative contributions of the individual channels to the coherent noise. The scalar parameter $c$ characterises the
level of the coherent noise in a given event.  
\item From this, the noise covariance matrix can be built as
\begin{equation}
B=\sigma^{2} \mathbf{1}+\sigma^{2}_{c} \boldsymbol{\alpha} \balpha^T, 
\label{eq:incohm}
\end{equation}
with $\langle u_{i}u_{j} \rangle =\sigma^{2}\delta_{ij}$ being the incoherent noise squared, $\mathbf{1}$ the unit matrix and $\sigma_{c}^{2}$ being the variance of the $c$-parameter introduced before.
\item The vector $\balpha$ is the eigenvector of $B$ with the largest eigenvalue given by $\omega_{1}=\sigma^{2}+\sigma_{c}^{2}$. In case of only one source of coherent noise, any other eigenvector orthogonal to $\balpha$ should have the eigenvalue $\sigma^2$.  In this model, the spectrum of eigenvalues is expected to be flat except for one eigenvalue from which the coherent noise can be derived. The Figure~\ref{fig:eigenval} shows the spectra of eigenvalues obtained for the four ASICs. 
The variance of the coherent noise $\sigma_{c}^{2}$ of the ASICs can be deduced from the largest eigenvalues and another one chosen from the flat parts of the spectra, which is reasonably fulfilled starting from $Rank=9$. The eigenvalues at that rank are defined as $\sigma^{2}$. The errors on the eigenvalues shown in the figure are derived according to the following plausibility consideration. According to~\cite{ieee2000} the eigenvalues are bounded by 
\begin{equation}
\lambda_{n} +\sigma^{2} b_{-}\leq \omega_n \leq \lambda_n+\sigma^{2} b_{+} 
\label{eq:bounds}
\end{equation} 
 
with $b_{\pm}=(1 \pm \sqrt{T/N} )^2$ where $T$ is the number of sources, here the ASIC channels, and $N$ is the number of events. The $\lambda_n$ are the true variances of the coherent noise where $\lambda_1 = \sigma_c^{2}$ in this analysis. The bounds span a range $\sigma^{2}(b_{+}-b_{-})$. For one source, i.e. $T=1$, this agrees with the statistical error of the variance multiplied by a factor  $\sqrt{2}$. Thus, to obtain the statistical error of the eigenvalues, the range of the bounds is calculated and divided by $\sqrt{2}$.  
 
After these considerations, it can be concluded that the eigenvalues for the two event types agree within statistical errors, which is particularly true for the largest one which carries the information on the coherent noise.

Figure~\ref{fig:eigenvec} shows the eigenvectors, normalised to unity, associated with the highest eigenvalue obtained in the same scan for signal and pedestal events. In some cases the eigenvectors feature the opposite sign. In that case also the reflected vector is given in the figure. This sign ambiguity of PCA is reported also in the literature, e.g.~\cite{bro2007}, and constitutes no hint of an inconsistency. The eigenvectors are thus in good agreement for signal and pedestal events.

A representation of the coherent channel noise can now be achieved by multiplying the variance, $\sigma_c^{2}$, with the component squared of the corresponding eigenvalue. The coherent noise is shown in Figure~\ref{fig:ccohn}.

It is clearly visible that for ASIC 1 and ASIC 3 the coherent noise is concentrated around the central channel numbers. The source of the coherent noise is not known but with a value of maximal 5\,$\rm{(ADC\,counts)}^ 2$, see Figure~\ref{fig:ccohn},  it is much smaller than the variance of the incoherent noise of about 20\,$\rm{(ADC\,counts)}^ 2$, see next step. There is no evidence that the observed coherent noise is different for signal and pedestal events. 

\begin{figure}
\centering
\includegraphics[width=0.78\textwidth]{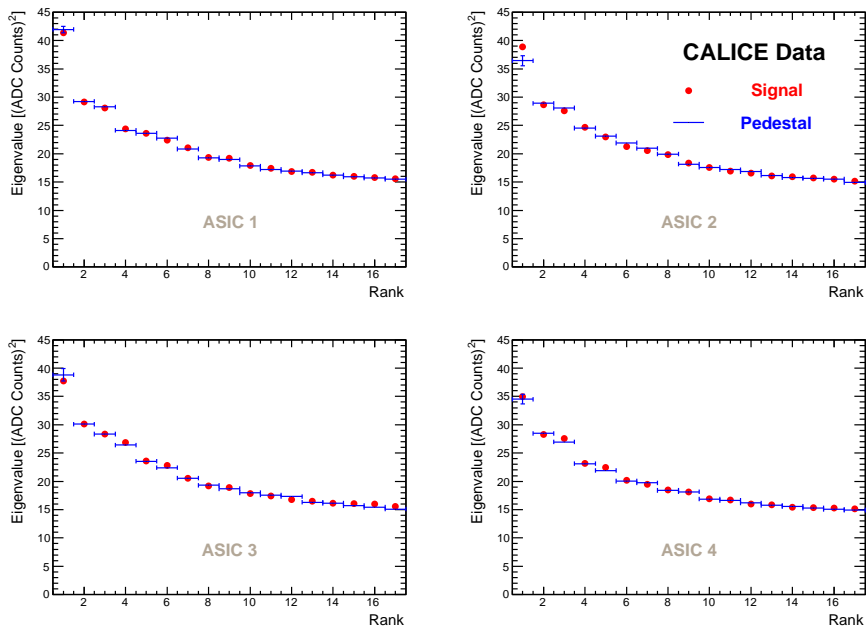}%
\caption{\sl Eigenvalues as obtained in a Principal Component Analysis of signal and pedestal events for nominal central impact
on the ASICs. The eigenvalues are ordered in decreasing order. The statistical error is given for the largest eigenvalue of the pedestal events.}
\label{fig:eigenval}
\end{figure}

\begin{figure}
\centering
\includegraphics[width=0.78\textwidth]{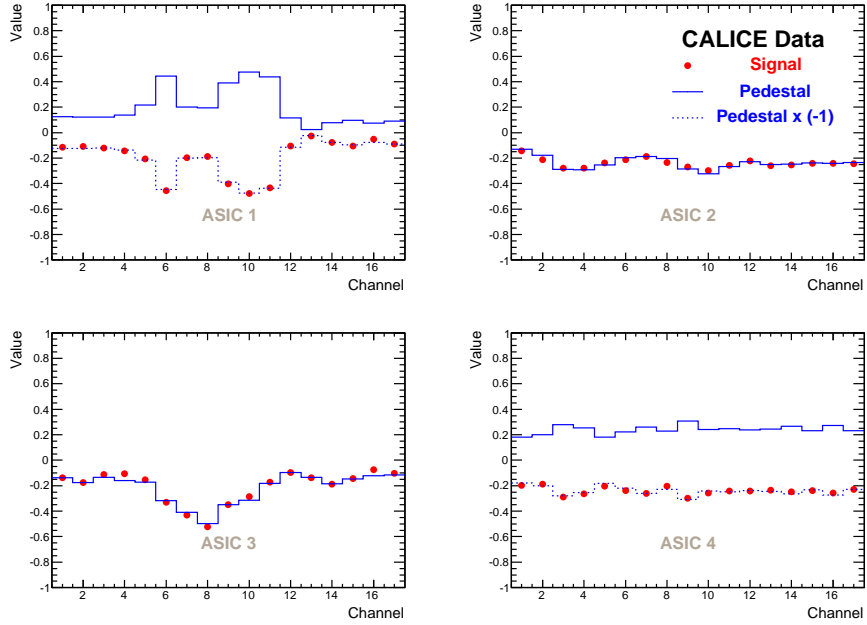}%
\caption{\sl Eigenvectors associated to the largest eigenvalues as obtained in a Principal Component Analysis of signal and pedestal events for nominal central impact on the ASICs. These eigenvectors indicate the location of coherent noise within the ASICs. In case of a sign flip between the eigenvectors for signal and pedestal events, the reflected eigenvector for pedestal events
is indicated, too.}
\label{fig:eigenvec}
\end{figure}

\begin{figure}
\centering
\includegraphics[width=0.83\textwidth]{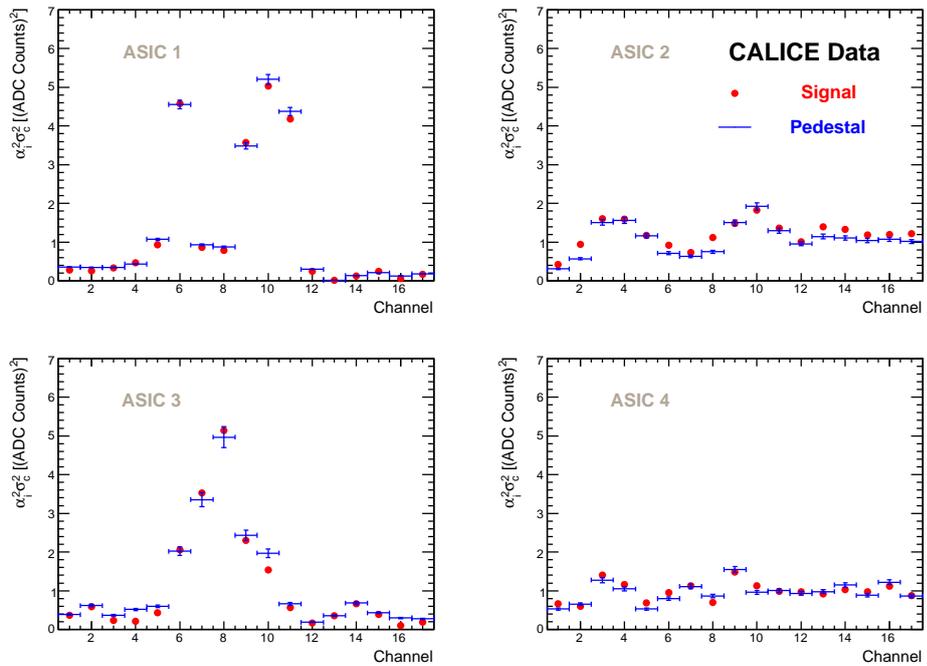}%
\caption{\sl Level of coherent noise in the four ASICs. }
\label{fig:ccohn}
\end{figure}

\item The incoherent noise per channel can be obtained from the deflated matrix 
\begin{equation}
B'=B-\sigma^{2}_{c} \balpha \balpha^T.
\end{equation}
In this matrix the off-diagonal elements are flat around a null value. The diagonal elements, however, can
be interpreted as the channel independent incoherent channel noise squared. The matrices obtained upon central impact on the ASICs in signal events are displayed in Figure~\ref{fig:inccohnm}. As expected, they feature dominant diagonal elements.

For confirmation, the diagonal elements are displayed separately in Figure~\ref{fig:inccohn}. As already mentioned above their
values are around 20\,$\rm{(ADC\,counts)}^ 2$ and channel independent.

\begin{figure}
\centering
\includegraphics[width=0.83\textwidth]{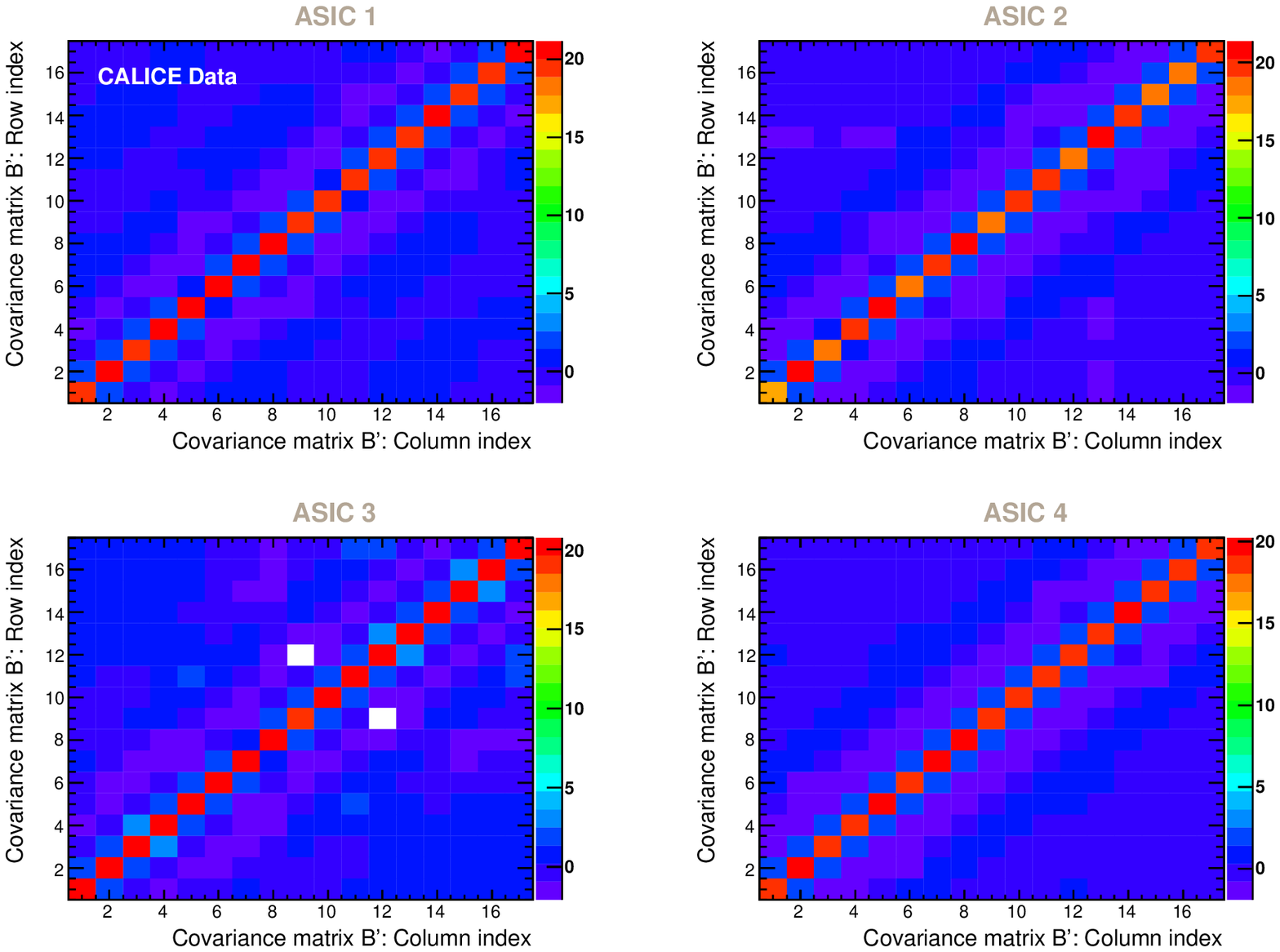}%
\caption{\sl Deflated matrix $B'$ for the four ASICs as obtained in a Principal Component Analysis for nominal central impact on the ASICs. Here, the matrices are shown for signal events. The colour represents the value of the matrix element in units of $\rm{(ADC\,counts)^2}$.}
\label{fig:inccohnm}
\end{figure}

\begin{figure}
\centering
\includegraphics[width=0.84\textwidth]{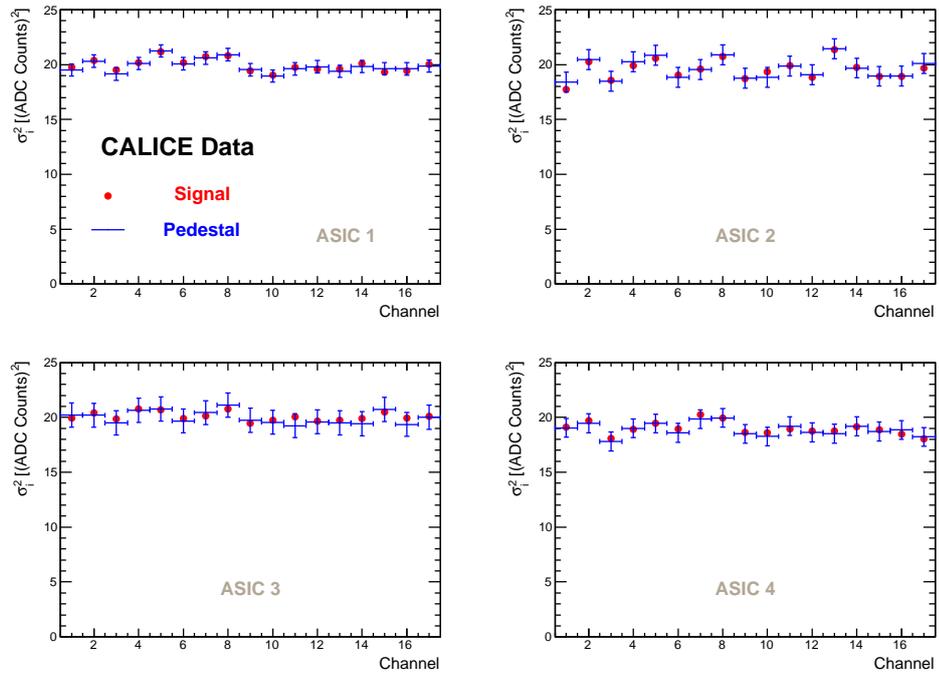}%
\caption{\sl Level of incoherent noise in the four ASICs. }
\label{fig:inccohn}
\end{figure}

In Section~\ref{sec:limits} the incoherent channel noise will be employed in the noise simulation of the ASICs.

\item Since all channels of an ASIC can be assumed to be equal, the variance $\sum u^2_{i}$ reaches a minimum. Thus, the $c$-parameter
can be estimated by requiring the quantity $\sum (b_i - \alpha_i)^2$ to be minimal. From this it follows that
\begin{equation} 
c=\balpha \cdot \mathbf{b}.
\end{equation}
Using this, the coherent noise could be estimated and subtracted on an event-by-event basis. In this analysis the knowledge of the
$c$-parameter together with other noise quantities will be exploited to simulate the noise of the ASICs. The $c$-parameter spectra for the four ASICs are given in Figure~\ref{fig:cpar}.

\begin{figure}
\centering
\includegraphics[width=0.84\textwidth]{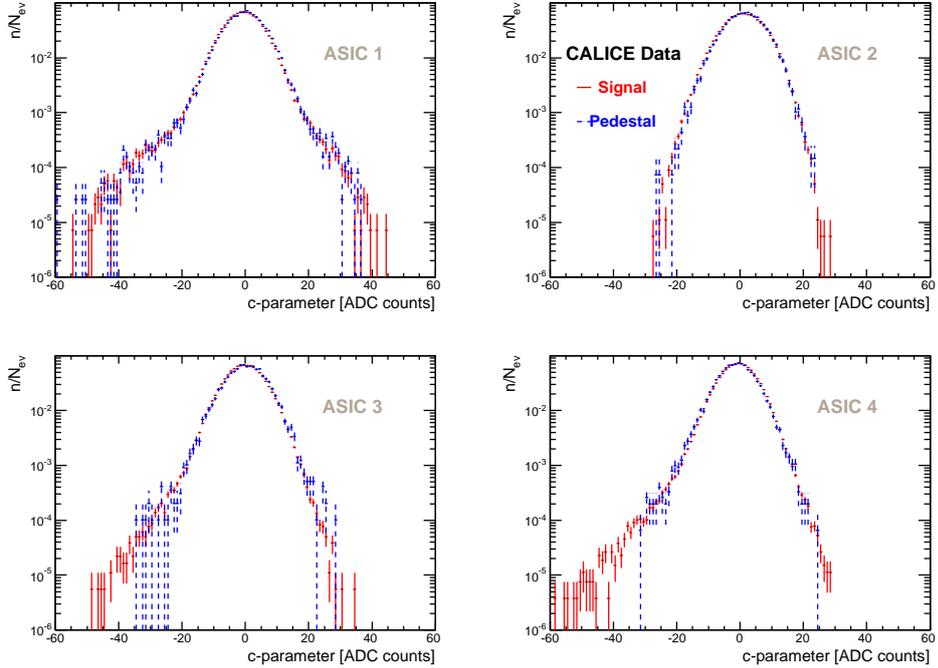}%
\caption{\sl Spectra of the $c$-parameter as obtained for nominal central impact on the four ASICs for signal and pedestal events. See text for the definition of the $c$-parameter. The histograms are normalised to unity.}
\label{fig:cpar}
\end{figure}
Again the spectra are very similar for signal and pedestal events. Since the statistics of the signal events are considerably larger than those of the pedestal events, the tails reach larger values.
\end{enumerate}

The PCA allows for the conclusion that the presence of shower particles has no significant influence on noise pattern of the ASICs. In addition, it indicates that, in a yet more quantitative study, the pedestal events can be used to model the noise pattern in the signal events. This will be done in the following section.

\section{Limits determination}
\label{sec:limits}
This section is dedicated to the determination of upper limits for having shower-induced fake hits above a given threshold. 
As a first result it can be reported that {\em no signal} above a MIP is observed in the signal and the pedestal events, such that the upper limit on the probability that shower particles induce a fake signal at the MIP level can be set to $6.7\cdot 10^{-7}$ at the 95\% confidence level. This number is derived from the run with the highest statistics listed in Table~\ref{tab:protocol}. In the remainder of this section, this result is extended towards smaller thresholds.
As it is rather expected that the shower induces fake signals towards small ADC counts, the threshold is varied between 15 and 30\,ADC counts, which corresponds to about $1/3$ to $2/3$ of the signal created by a MIP.
This covers the region of noise cuts studied in~\cite{calice2} and allows for the investigation of the influence of the particle shower towards smallest ADC values. Values smaller than 15\,ADC counts have been discarded as these show a large sensitivity to the residual pedestal instabilities.

\subsection*{Limits on signals in the presence of background}

The comparatively small ADC values require the determination of the upper limits in the presence of  background given by the intrinsic noise of the ASICs. In this case, the 
Poissonian probability density function $f'$ for observing $n$ events based on a sample statistics $\mathrm{k}$ is given by~\cite{brandt}:
\begin{equation}   
f'(n; \lambda_S+\lambda_B)=f(n; \lambda_{S}+\lambda_{B})/\sum^{\mathrm{k}}_{n_{B}=0}f(n_B; \lambda_B). 
\label{eq:poisspdf}
\end{equation}
Here, $\lambda_S$ and $\lambda_B$ are the Poisonnian parameters for {\em S}ignal and {\em B}ackground, respectively. 
The probability density function $f(n; \lambda_{S}+\lambda_{B})$ is the sum of the independent Poissonian distributions for
signal and background to the Poissonian parameter $\lambda=\lambda_S+\lambda_B$. The probability distribution function in the 
denominator of Equation~\ref{eq:poisspdf} ensures that $f'$ is normalised to 1 for background only events. The sum runs over the possible number of background 
events, $n_B$, up to the sample statistics $\mathrm{k}$. The probability distribution function to the probability density in Equation~\ref{eq:poisspdf} reads 
\begin{equation}   
F'(\mathrm{k}; \lambda_S+\lambda_B)=\sum^{\mathrm{k-1}}_{n=0} f'(n; \lambda_{S}+\lambda_{B})=P(\ell<\mathrm{k}),
\label{eq:cumulpdf}
\end{equation}
where $P(\ell<\mathrm{k})$ is the probability to observe any number $\ell<\mathrm{k}$. The upper limit $\lambda^{(up)}_S$ for signal events at the confidence level $\beta=1-\alpha$ can then be obtained from
\begin{equation}   
\alpha=F'(\mathrm{k+1}; \lambda^{(up)}_S+\lambda_B)
\label{eq:uplim}
\end{equation}
in case the background is known. 

As there is no indication that the high-energy showers influence the ASIC response, the limits will be derived for those four measurement points in which the beam was incident on the nominal centre of one of the ASICs. The background expectation will be obtained from simulated events.

\subsection{Noise simulation}

The simulation of the noise starts out from the noise vector given in Equation~\ref{eq:noisevec}. The incoherent noise is thus 
simulated using a Gaussian $G(x_{m}, \sigma_i)$ with the $\sigma_i$ of the individual ASIC-channels read off from the 
matrices given in  Equation~\ref{eq:incohm}. The mean $x_m$ of the Gaussian is given by the mean measured in the pedestal events in a given run. The part covering the coherent noise is realised by simulating the $c$-parameter spectrum and by multiplying this spectrum with the corresponding component $\alpha_i$ of the eigenvector of a given ASIC. Figure~\ref{fig:cpar} illustrates that the $c$-parameter spectrum cannot be approximated by a simple Gaussian. Rather, it is simulated using an adaptive kernel estimation introduced in~\cite{cranmer}. Here, the kernel estimation which corresponds to the implementation in the RooFit package is  employed. The formula used to simulate the noise spectrum $S'_{i}$ for a channel $i$ thus reads
\begin{equation}
S'_{i}= G(x_{m}, \sigma_i)+(sign)K(c)\alpha_i.
\label{eq:sim}
\end{equation}

The symbol $K(c)$ describes the kernel estimation introduced before. The $sign$ is given by the scalar product of  the eigenvectors obtained for the signal events and pedestal events. 
The first aim of the simulation is to reproduce the measured pedestal spectra in this paper. Here and in the following it is ensured that the number of simulated events is at least 2.5 times larger than the number of measured events. Thus the statistical error of the simulation is smaller than that of the data. In order to obtain a maximal level of agreement between the simulated and the measured pedestal spectra, two free parameters are added to the Equation~\ref{eq:sim} leading to:

\begin{equation}
S_{i}= G(x_{m} - p_{m}, \sqrt{\sigma_i^2 - p_{\sigma}})+(sign)K(c)\alpha_i.
\label{eq:simtun}
\end{equation}

\begin{figure}
\centering
\includegraphics[width=0.83\textwidth]{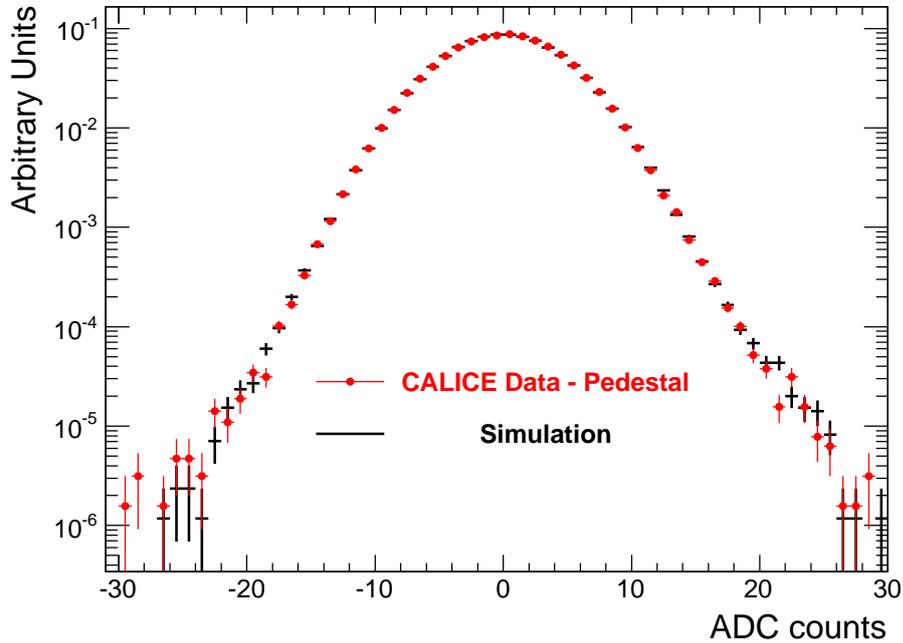}%
\caption{\sl Comparison between the measured and simulated pedestal spectrum at the example of  ASIC 1 in Scan 1  and measurement point  3. The details of the simulation are explained in the text.}
\label{fig:peddatsim}
\end{figure}

\begin{figure}
\centering
\includegraphics[width=0.8\textwidth]{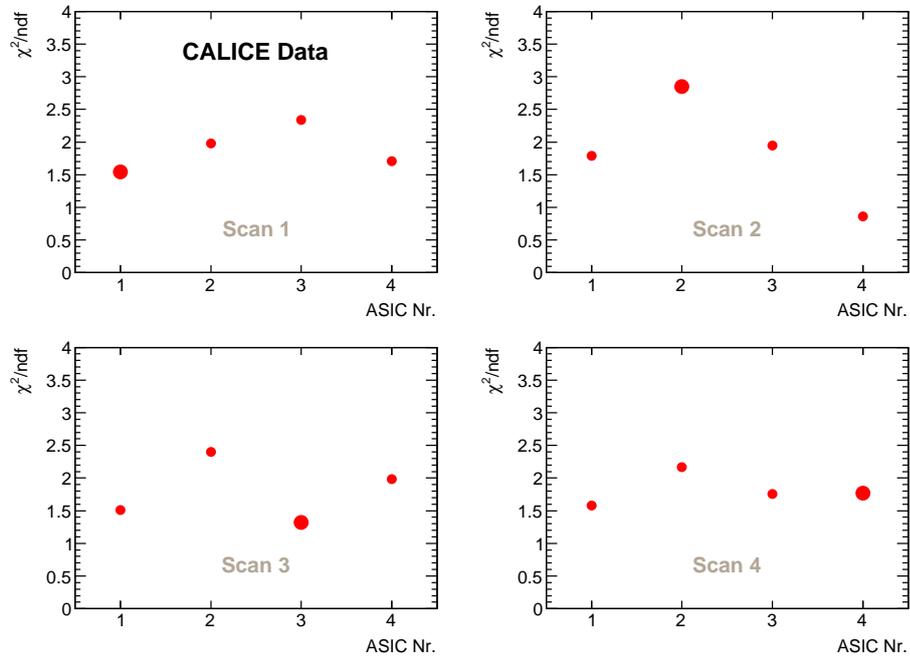}%
\caption{\sl Resulting $\chi^{2}/ndf$ of the comparison between the measured and simulated pedestal spectra. The comparisons are made for the measurement points 3 in the scans, see Table~\ref{tab:protocol}. The ASIC which is actually exposed to the beam is indicated by a larger symbol.The ASIC number is given on the $x$-axis of the graph.}
\label{fig:chi2comp}
\end{figure}

These free parameters are used to account for residual off-diagonal elements in the matrices of Equation~\ref{eq:incohm}. 
In addition, they account for imperfections caused by the loss of information in using only the eigenvector with the largest eigenvalue. The free parameters are tuned until a minimum $\chi^{2}/ndf$ is obtained upon comparing the spectra of the pedestal events with the simulated ones. The range of values of the free parameters are $p_{m}=[-0.33,0.25]$ and $p_{\sigma}=[1.1,2.0]$. A comparison between the measured pedestal spectrum of ASIC 1 for Scan 1 
and measurement point  3 is given in Figure~\ref{fig:peddatsim}. An excellent agreement between data and simulation is achieved. The resulting $\chi^{2}/ndf$ as a function of the ASIC  number for all runs with central impact on one of the ASICs is shown in Figure~\ref{fig:chi2comp}. 

\begin{figure}
\centering
\includegraphics[width=0.8\textwidth]{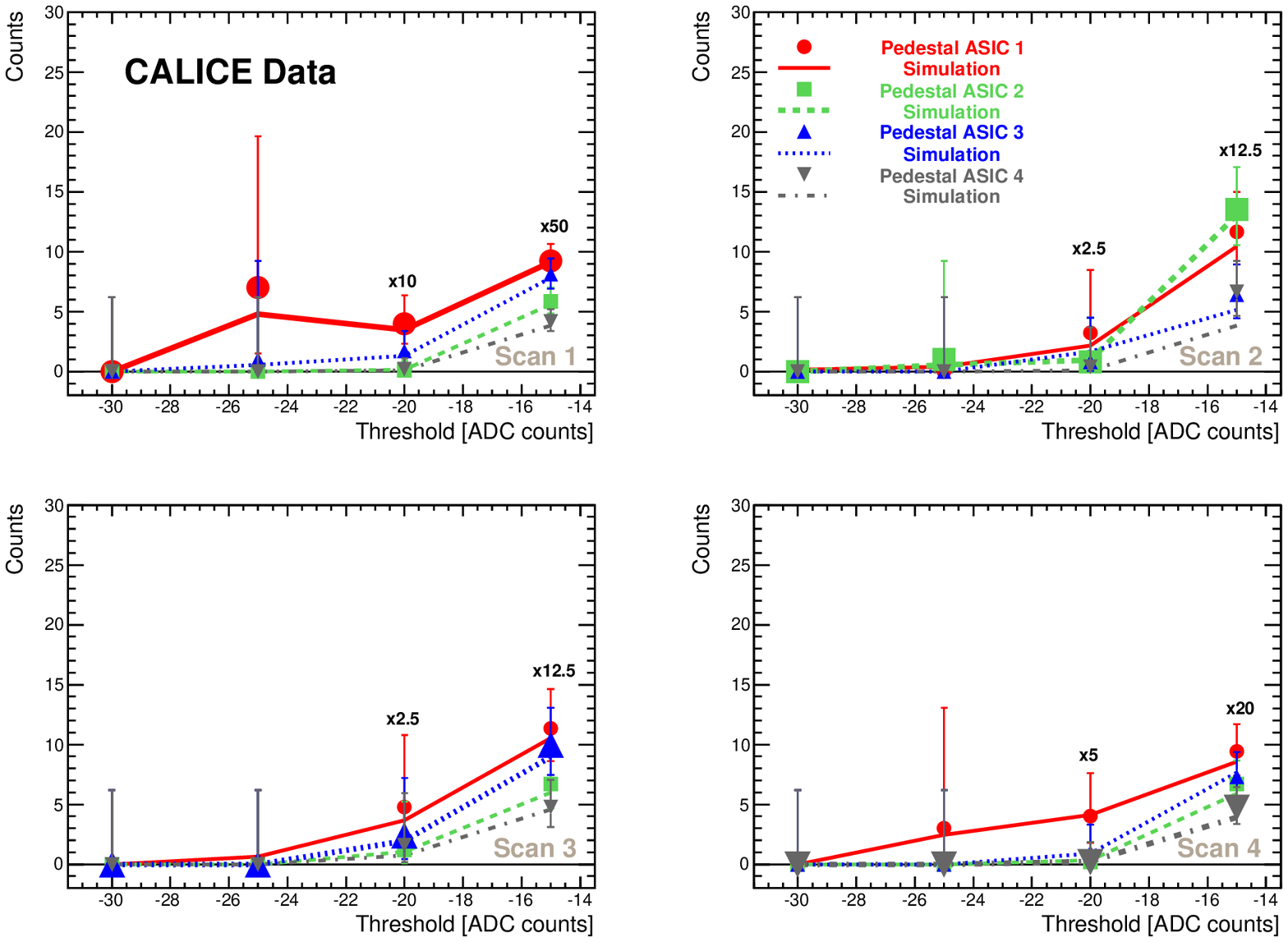}%
\caption{\sl The measured counts below a threshold in pedestal events are compared with those obtained from simulation. The results given in this figure are for negative ADC counts. The error bars on the data points correspond to the 97.3\% confidence interval.  The comparisons are made for the measurement points 3 in the scans, see Table~\ref{tab:protocol}. The ASIC which is exposed to the beam is indicated by a larger symbol. For small absolute thresholds the number of counts have been downscaled by the factors indicated in the figure.}
\label{fig:negped}
\end{figure}

\begin{figure}
\centering
\includegraphics[width=0.8\textwidth]{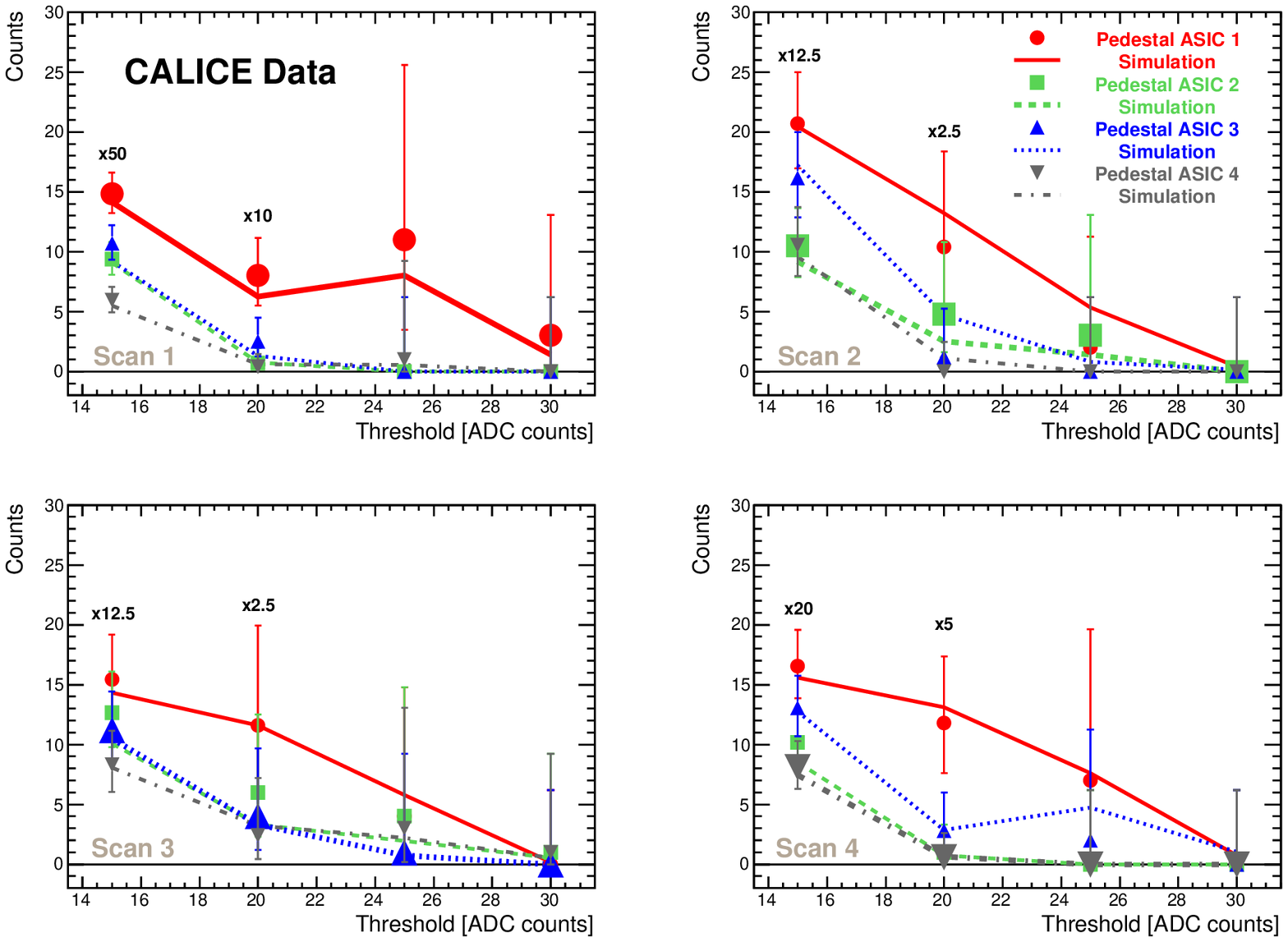}%
\caption{\sl Same as Figure~\ref{fig:negped} but for positive hits.}
\label{fig:posped}
\end{figure}

Inspired by the work presented in~\cite{bogdan}, the number of hits in the simulated spectra are subject to a final correction. For each bin the inverse error function is calculated according to
\begin{equation}
e(\Delta x, \sigma)=\Bigg[1-erf\Bigg(\frac{\Delta x}{2\sqrt{2}\sigma_b}\Bigg)\Bigg]\cdot \Delta x.
\end{equation}
Here, $\Delta x$ is the difference between simulation and pedestal events and $\sigma_b$ is the statistical uncertainty of the data in that bin.
This correction aims to balance out residual imperfections of the simulation without being too sensitive to statistical fluctuations appearing in the tails of the spectra of the pedestal events.
After this final correction, the number of hits above a given threshold in pedestal events is compared for data and simulation. The comparisons are made separately for negative and positive ADC counts and are shown in Figures~\ref{fig:negped} and~\ref{fig:posped}.
The error on the data is given by the 97.3\% confidence interval around the measured number of hits. Towards large numbers of hits, this corresponds to the $3\sigma$ confidence interval. This interval is chosen since due to the complexity of the noise spectra not all potential effects are incorporated in the simulation. Systematic uncertainties of the noise modelling can be estimated from the spread of the simulated distributions of the four ASICs, which should in principle be identical. 
Data and simulation agree within the chosen confidence limit where the agreement is better for negative ADC counts.
The agreement achieved validates the model for making reliable predictions of the expected number of events in signal events.


\begin{figure}
\centering
\includegraphics[width=0.8\textwidth]{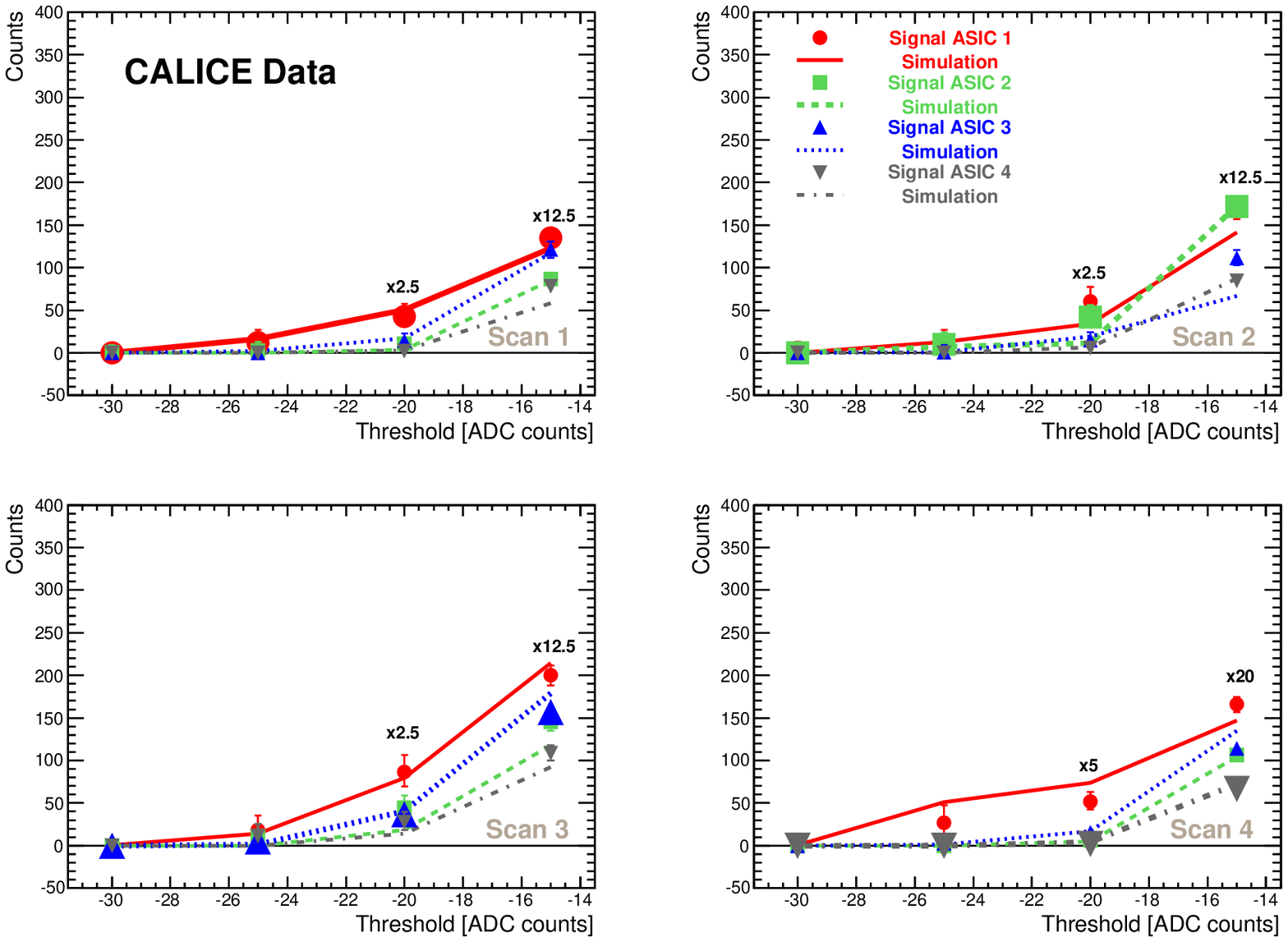}%
\caption{\sl The measured counts below a threshold in signal events are compared with simulation. The results given in this figure are for negative ADC counts. The error bars on the data points correspond to the 97.3\% confidence interval.  The comparisons are made for the measurement points 3 in the scans, see Table~\ref{tab:protocol}. The ASIC that is exposed to the beam is indicated by a larger symbol. For small absolute thresholds the number of counts have been downscaled by the factors indicated in the figure.}
\label{fig:negsig}
\end{figure}

\begin{figure}[!h]
\centering
\includegraphics[width=0.8\textwidth]{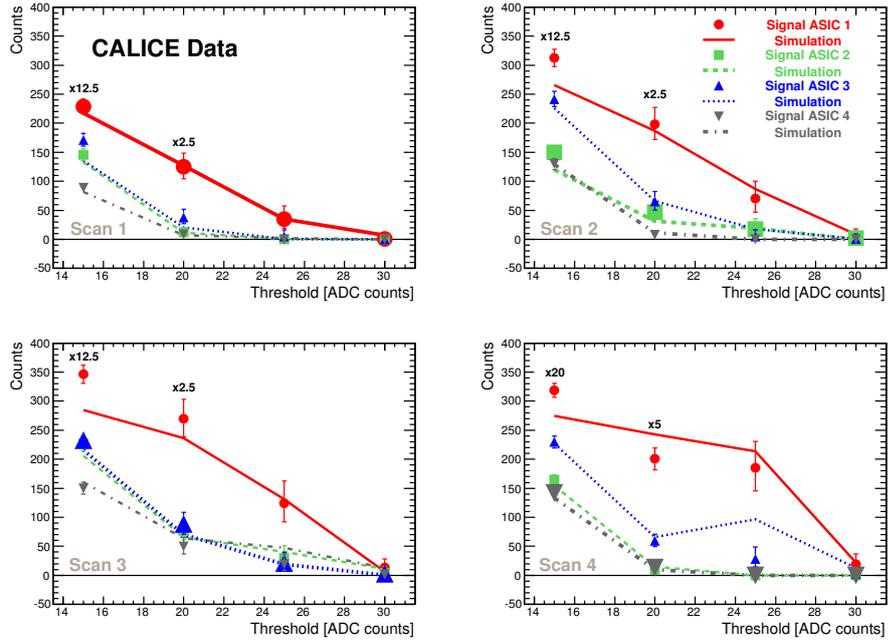}%
\caption{\sl Same as Figure~\ref{fig:negsig} but for positive hits.}
\label{fig:possig}
\end{figure}

The comparison of the simulated spectra with those measured in the signal events is made in Figures~\ref{fig:negsig} and~\ref{fig:possig}. Data and simulation are still compatible. For most of the data points the agreement 
is within the chosen confidence interval. The deviations may not be attributed to the actual beam exposure since they occur in their
majority for ASICs outside the actual electromagnetic shower. The agreement is worse towards smaller ADC values 
while towards large ADC values data and simulation agree within the chosen limits. The former discussion indicates that there is no measurable influence of the beam on the ASIC  response. Therefore, in the following the upper limits on the frequency of fake hits are determined.

\clearpage

\begin{figure}
\centering
\includegraphics[width=0.8\textwidth]{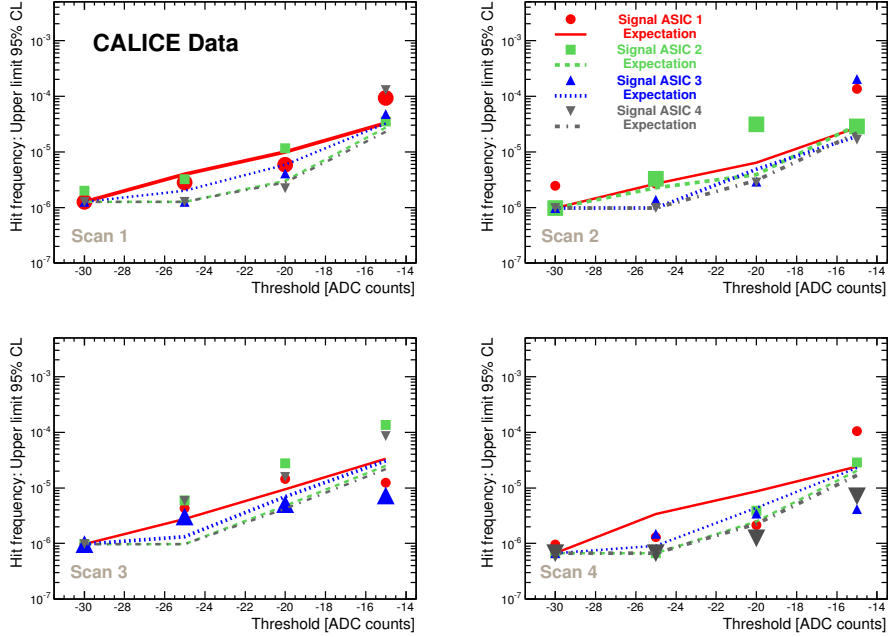}%
\caption{\sl Upper limits on frequencies at the 95\% confidence level of fake hits compared with those expected from the pure pedestal events.  The limits are given as a function of a threshold for negative values of the ADC counts. The limits are shown for the measurement points 3 in the scans, see Table~\ref{tab:protocol}. The ASIC which is exposed to the beam is indicated by a larger symbol.}
\label{fig:neglim}
\end{figure}

\begin{figure}
\centering
\includegraphics[width=0.8\textwidth]{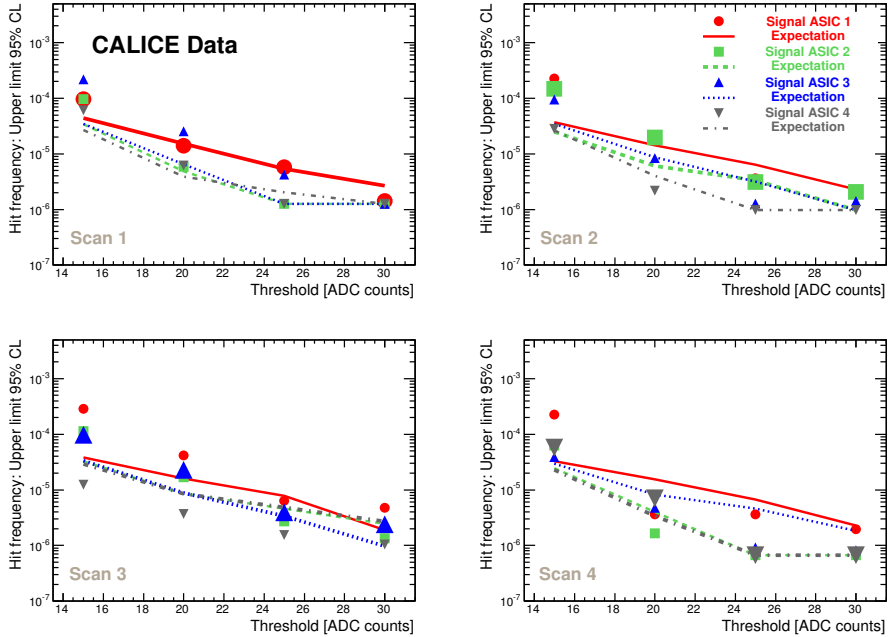}%
\caption{\sl Same as Figure~\ref{fig:neglim} but for positive hits.}
\label{fig:poslim}
\end{figure}

In application of Equations~\ref{eq:poisspdf},~\ref{eq:cumulpdf} and~\ref{eq:uplim}, the sample statistics $k$ is given by the
number of hits above a given threshold and $\lambda_{B}$, i.e. the number of expected hits, is obtained from simulation.
From this  $\lambda^{(up)}_S$ is derived using a computer program available in~\cite{brandt}. Finally, $\lambda^{(up)}_S$ is divided by the total number of hits to calculate the frequency of fake hits. 
The upper limits at the $\beta=95\%$ confidence level  on the frequency of fake hits are shown in Figures~\ref{fig:neglim} and~\ref{fig:poslim} separately for positive and negative ADC counts. 

For each scan, the upper limits are determined for each of the four ASICs in order to compare the behaviour of an ASIC exposed to the electron showers with those not exposed to the showers.  It may be seen that the determined upper limits are always smaller than $5\cdot10^{-4}$ for the smallest threshold value and smaller than $10^{-5}$ for the highest threshold value. The observed dependency on the threshold is the same whether or not the ASICs are exposed to the particle showers. The upper limits are compared with sensitivity limits, full lines in Figures~\ref{fig:neglim} and~\ref{fig:poslim}, obtained when the frequencies observed in data are replaced by those expected from the simulation. In particular towards large threshold values the derived limits agree well with the expectation. Deviations from the expectation are observed both for ASICs exposed to showers and for those outside the showers.

This observation renders it unlikely that there is an influence of the beam on the measured signal. This is in particular true for threshold values relevant for physics analysis in which typically values below 25 ADC counts are discarded. Remaining deviations towards small thresholds from the expectation can be attributed to the influence by other parts of the experimental set-up which are present when there is a large activity in the detector. 
These interspersed signals are not taken into account in the simulation. 


\section{Conclusion and outlook}

A series of test runs has been performed and analysed in order to prove the feasibility of having embedded readout electronics 
for a calorimeter proposed for a future lepton collider. A detailed analysis of noise spectra of the ASICs exposed to high-energy electron beams has revealed no evidence that the noise pattern is altered under the influence of the electromagnetic showers. 
The probability to have fake signals above the MIP level is estimated to be smaller than $6.7\cdot 10^{-7}$. The probability for a fake signal 
is less than $10^{-5}$ for a threshold of 2/3 of a MIP. For an event of the type $\epem \rightarrow \ttbar$ at $\roots=500\,\GeV$ at the lepton collider about 2500 cells of dimension $1\times1\,\mathrm{cm^2}$ are expected to carry a signal above noise level which is typically defined to be (60-70)\% of a MIP. The results presented in this article have revealed no problems for the design of embedded readout electronics for a detector for a lepton collider.
It is furthermore unlikely that the residual deviations between the observed number of hits and those expected from normal noise fluctuations can be attributed to the influence of the beam but rather to an imperfect modelling of the noise spectra for signal events. In this sense, the presented results constitute a conservative upper limit.

Currently, the CALICE collaboration is about to construct a technological prototype~\cite{eudet}. In contrast to the physics prototype, this technological prototype will have the readout electronics embedded by design. The ASICs employed therein are a straightforward further development of those of the physics prototype~\cite{dlt09} as described in this paper. The technological prototype in general and the ASICs in particular are close to the design currently envisaged for the International Linear Collider which is currently the most advanced proposal for a future lepton collider.
A series of tests as described in this article will have to be repeated for this prototype as the electronics are more challenging than 
the one employed in the physics prototype with respect to compactness and requirements of power saving. Upon repetition of the test a considerably larger amount of interleaved  pedestal events will have to be recorded. Future tests should also be conducted with heavily ionising particles up to the point at which radiation effects become apparent. With cross talk effects further reduced, such a research program will allow for the establishment of a complete picture of the feasibility of embedded electronics in radiation environments. 

\section*{Acknowledgements}
We would like to thank the technicians and the engineers who
contributed to the design and construction of the prototypes. CALICE conducts 
test beams at CERN, DESY and FNAL and we gratefully acknowledge the managements of these laboratories 
for their support and hospitality, and their accelerator staff for the reliable and efficient
beam operation. 
We would like to thank the HEP group of the University of
Tsukuba for the loan of drift chambers for the DESY test beam.
This work was supported within the 'Quarks and Leptons' programme of the CNRS/IN2P3, France;
by the Bundesministerium f\"{u}r Bildung und Forschung, Germany;
by the  the DFG cluster of excellence `Origin and Structure of the Universe' of Germany; 
by the Helmholtz-Nachwuchsgruppen grant VH-NG-206;
by the BMBF, grant no. 05HS6VH1;
by the Alexander von Humboldt Foundation (Research Award IV, RUS1066839 GSA);
by MICINN and CPAN, Spain;
by the US Department of Energy and the US National Science Foundation;
by the Ministry of Education, Youth and Sports of the Czech Republic
under the projects AV0 Z3407391, AV0 Z10100502, LC527  and LA09042  and by the
Grant Agency of the Czech Republic under the project 202/05/0653;  
and by the Science and Technology Facilities Council, UK.

\newpage
\section*{Appendix - Mean and RMS of signal and pedestal events}

\begin{figure}[ht]
\begin{center}
\includegraphics[width=0.9\textwidth]{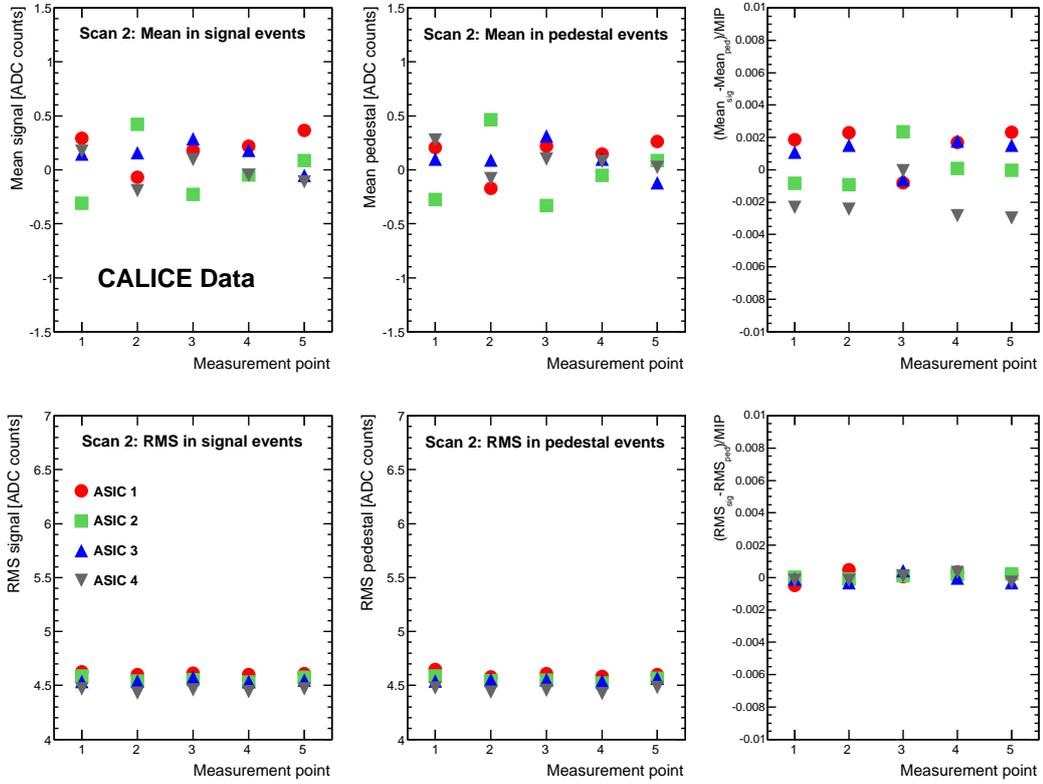}
\caption{\sl Mean and RMS for signal events and pedestal events. The very right part shows the corresponding differences normalised to the value of a MIP assumed to be 45 ADC counts. The figure displays the result for Scan 2 in which ASIC 2 is scanned. As a cross-check the results for all ASICs are shown.}
\label{fig:cscan2}
\end{center}
\end{figure}
 
\begin{figure}[ht]
\begin{center}
\includegraphics[width=0.9\textwidth]{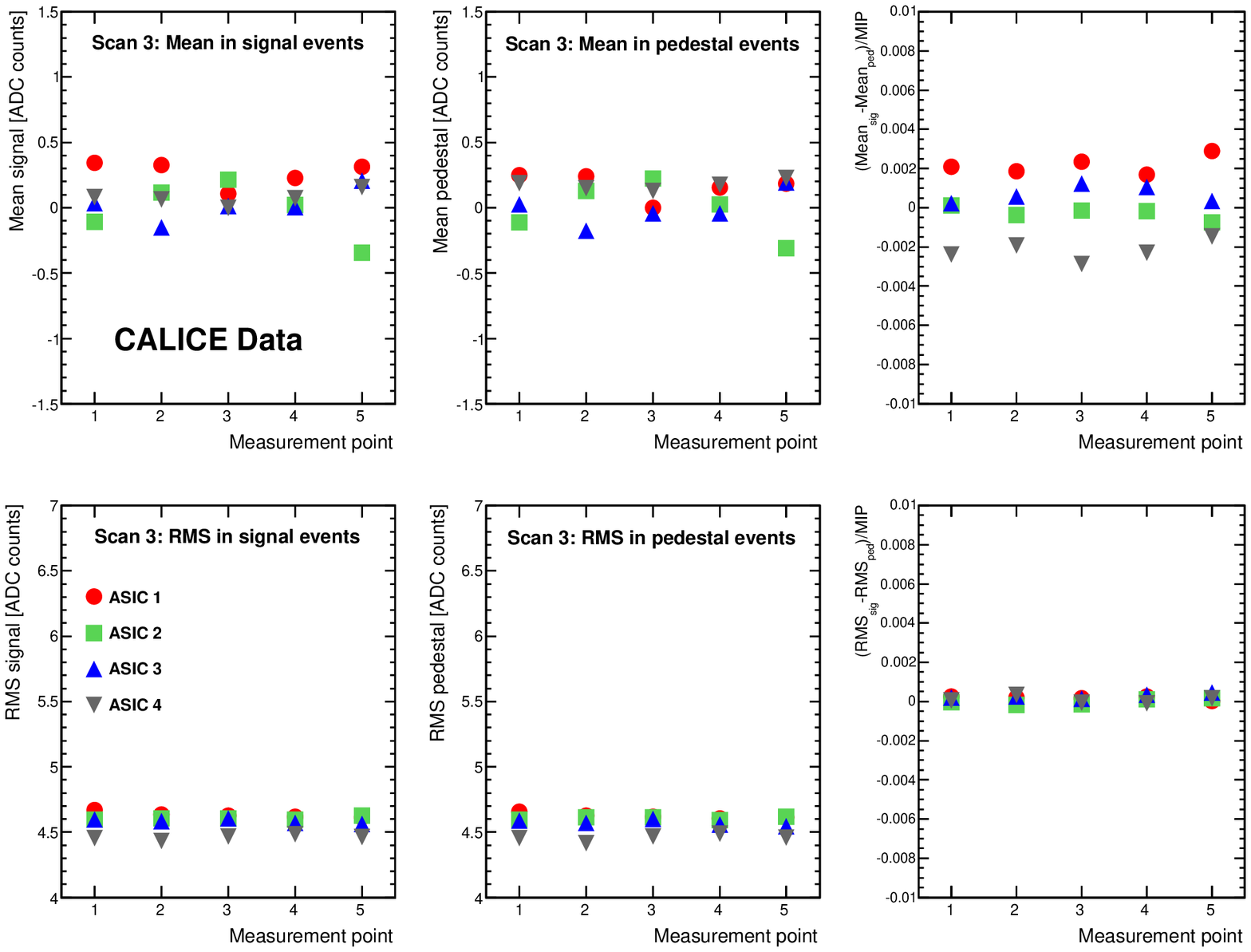}
\caption{\sl Mean and RMS for signal events and pedestal events. The very right part shows the corresponding differences normalised to the value of a MIP assumed to be 45 ADC counts. The figure displays the result for Scan 3 in which ASIC 3 is scanned. As a cross-check the results for all ASICs are shown.}
\label{fig:cscan3}
\end{center}
\end{figure}

\begin{figure}[ht]
\begin{center}
\includegraphics[width=0.9\textwidth]{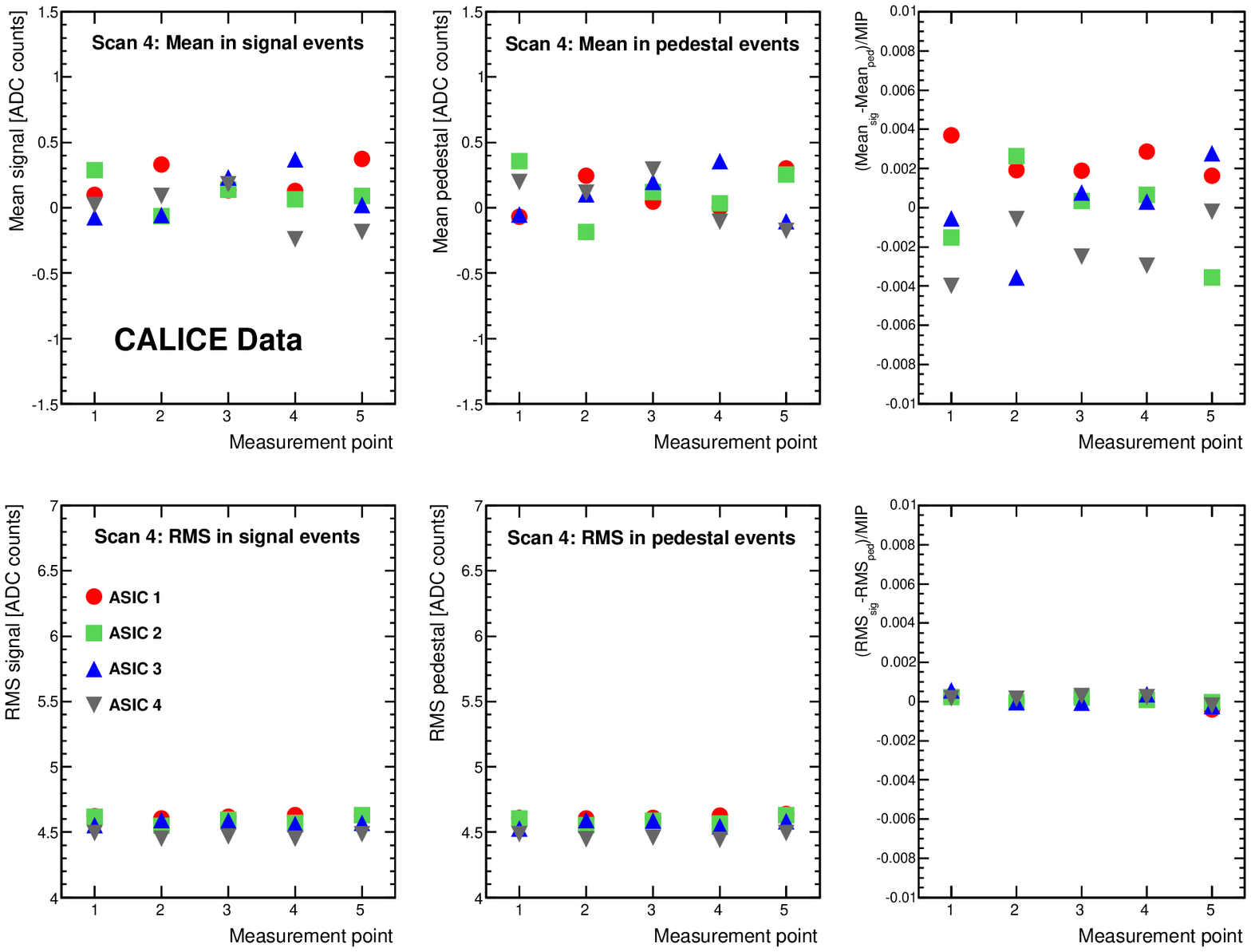}
\caption{\sl Mean and RMS for signal events and pedestal events. The very right part shows the corresponding differences normalised to the value of a MIP assumed to be 45 ADC counts. The figure displays the result for Scan 4 in which ASIC 4 is scanned. As a cross-check the results for all ASICs are shown.}
\label{fig:cscan4}
\end{center}
\end{figure}
\clearpage

\newpage
\begin{footnotesize}



%

\end{footnotesize}

\end{document}

%% file: authors_inc.tex
\setcounter{page}{1}

\begin{center}
C.\,Adloff
\begin{footnotesize}\\
\it Laboratoire d'Annecy-le-Vieux de Physique des Particules, Universit\'{e} de Savoie,
CNRS/IN2P3,
9 Chemin de Bellevue BP110, F-74941 Annecy-le-Vieux CEDEX, France

\end{footnotesize}
K.\,Francis,
J.\,Repond, 
J.\,Smith\endnote{Also at University of Texas, Arlington},
D.\,Trojand\endnote{Also at McGill University, Department of Physics, 3600 University Street, Montr\'eal, Qu\'ebec, Canada, H3A 2T8 },
L.\,Xia 
\\ 
\begin{footnotesize}
\it Argonne National Laboratory, 9700 S.\ Cass Avenue, Argonne, IL 60439-4815, USA
\end{footnotesize}

E.\,Baldolemar, 
J.\,Li\endnote{Deceased}, 
S.\,T.\,Park, 
M.\,Sosebee, 
A.\,P.\,White, 
J.\,Yu
\\ 
\begin{footnotesize}
\it Department of Physics, SH108, University of Texas, Arlington, TX 76019, USA
\end{footnotesize}

Y.\,Mikami, 
N.\,K.\,Watson
\\ 
\begin{footnotesize}
\it University of Birmingham, School of Physics and Astronomy, Edgbaston, Birmingham B15 2TT, UK
\end{footnotesize}

G.\,Mavromanolakis \endnote{Now at CERN}, 
M.\,A.\,Thomson, 
D.\,R.\,Ward, 
W.\,Yan\endnote{Now at Dept.\, of Modern Physics, Univ. of Science and Technology of China, 96 Jinzhai Road, Hefei, Anhui, 230026, P.\, R.\, China}
\\
\begin{footnotesize}
\it University of Cambridge, Cavendish Laboratory, J J Thomson Avenue, CB3 0HE, UK
\end{footnotesize}

D.\,Benchekroun, 
A.\,Hoummada, 
Y.\,Khoulaki
\\ 
\begin{footnotesize}
\it R\'{e}seau Universitaire de Physique des Hautes Energies (RUPHE),
Universit\'{e} Hassan II A\"{\i}n Chock, Facult\'{e} des sciences.\, B.P. 5366 Maarif, Casablanca, Morocco
\end{footnotesize}

M.\,Benyamna, 
C.\,C\^{a}rloganu, 
F.\,Fehr, 
P.\,Gay, 
S.\,Manen, 
L.\,Royer
\\ 
\begin{footnotesize}
\it Clermont Universit\'e, Universit\'e Blaise Pascal, CNRS/IN2P3, LPC, BP
10448, F-63000 Clermont-Ferrand, France
\end{footnotesize}

G.\,C.\,Blazey,
A.\,Dyshkant, 
V.\,Zutshi
\\ 
\begin{footnotesize}
\it NICADD, Northern  Illinois University, Department of Physics,
DeKalb, IL 60115, USA
\end{footnotesize}

J.\,-Y.\,Hostachy, 
L.\,Morin
\\ 
\begin{footnotesize}
\it Laboratoire de Physique Subatomique et de Cosmologie - Universit\'{e} Joseph Fourier Grenoble 1 - CNRS/IN2P3 - Institut Polytechnique de Grenoble,
53, rue des Martyrs,
38026 Grenoble CEDEX, France
\end{footnotesize}

U.\,Cornett, 
D.\,David, 
R.\,Fabbri, 
G.\,Falley, 
K.\,Gadow, 
E.\,Garutti,
P.\,G\"{o}ttlicher, 
C.\,G\"{u}nter,
S.\,Karstensen, 
F.\,Krivan,
A.\,-I.\,Lucaci-Timoce\endnotemark[4], 
S.\,Lu, 
B.\,Lutz, 
I.\,Marchesini, 
N.\,Meyer,
S.\,Morozov, 
V.\,Morgunov\endnote{On leave from ITEP}, 
M.\,Reinecke, 
F.\,Sefkow, 
P.\,Smirnov,
M.\,Terwort,
A.\,Vargas-Trevino, 
N.\,Wattimena, 
O.\,Wendt
\\ 
\begin{footnotesize}
\it DESY, Notkestrasse 85,
D-22603 Hamburg, Germany
\end{footnotesize}

N.\,Feege, 
J.\,Haller, 
S.\,Richter, 
J.\,Samson
\\ 
\begin{footnotesize}
\it Univ. Hamburg,
Physics Department,
Institut f\"ur Experimentalphysik,
Luruper Chaussee 149,
22761 Hamburg, Germany
\end{footnotesize}

P.\,Eckert,
A.\,Kaplan,
 H.\,-Ch.\,Schultz-Coulon,
 W.\,Shen,
 R.\,Stamen,
 A.\,Tadday
\\ 
\begin{footnotesize}
\it
 University of Heidelberg, Fakult\"at fur Physik und Astronomie,
Albert-Ueberle Str. 3-5 2.OG Ost, D-69120 Heidelberg, Germany
\end{footnotesize}

B.\,Bilki, E.\,Norbeck, 
Y.\,Onel
\\ 
\begin{footnotesize}
\it
University of Iowa, Dept. of Physics and Astronomy,
203 Van Allen Hall, Iowa City, IA 52242-1479, USA
\end{footnotesize}

K.\,Kawagoe,  
S.\,Uozumi\endnote{Now at Kyungpook Nation University, 1370 Sankyuk-dong, Buk-go Daegu 701-701, Korea}
\\ 
\begin{footnotesize}
\it  Department of Physics, Kobe University, Kobe, 657-8501, Japan
\end{footnotesize}

P.\,D.\,Dauncey, 
A.\,-M.\,Magnan
\\ 
\begin{footnotesize}
\it
Imperial College London, Blackett Laboratory,
Department of Physics,
Prince Consort Road,
London SW7 2AZ, UK 
\end{footnotesize}

V.\,Bartsch\endnote{Now at University of Sussex, Physics and
  Astronomy Department, Brighton, Sussex, BN1 9QH, UK}
\\ 
\begin{footnotesize}
\it Department of Physics and Astronomy, University College London,
Gower Street,
London WC1E 6BT, UK
\end{footnotesize}

F.\,Salvatore\endnotemark[8]
\\ 
\begin{footnotesize}
\it
Royal Holloway University of London,
Dept. of Physics,
Egham, Surrey TW20 0EX, UK
\end{footnotesize}

I.\,Laktineh
\\ 
\begin{footnotesize}
\it
Universit\'{e} de Lyon, Universit\'{e} de Lyon 1, 
CNRS/IN2P3, IPNL 4 rue E Fermi 69622,
Villeurbanne CEDEX, France
\end{footnotesize}

E.\,Calvo~Alamillo, 
M.-C.\, Fouz, 
J.\,Puerta-Pelayo 
\\ 
\begin{footnotesize}
\it
CIEMAT, Centro de Investigaciones Energeticas, Medioambientales y Tecnologicas, Madrid, Spain 
\end{footnotesize}

\setcounter{footnote}{1}

A.\,Frey\endnote{Now at University of G\"{o}ttingen, II. Physikalisches Institut
Friedrich-Hund-Platz 1, 37077 G\"ottingen, Germany}, 
C.\,Kiesling,
F.\,Simon
\\ 
\begin{footnotesize}
\it
Max Planck Inst. f\"ur Physik,
F\"ohringer Ring 6,
D-80805 Munich, Germany
\end{footnotesize}

J.\,Bonis, 
B.\,Bouquet,    
S.\,Callier, 
P.\,Cornebise, 
Ph.\,Doublet,
F.\,Dulucq, 
M.\,Faucci Giannelli, 
J.\,Fleury,
H.\,Li\endnote{Now at LPSC Grenoble},  
G.\,Martin-Chassard, 
F.\,Richard, 
Ch.\,de la Taille, 
R.\,P\"{o}schl, 
L.\,Raux,  
N.\,Seguin-Moreau, 
F.\,Wicek
\\ 
\begin{footnotesize}
\it 
Laboratoire de l'Acc\'{e}l\'{e}rateur Lin\'{e}aire, Centre Scientifique d'Orsay, Universit\'{e} de Paris-Sud XI, CNRS/IN2P3, BP 34, B\^atiment 200, F-91898 Orsay Cedex, France
\end{footnotesize}

M.\,Anduze,
V.\,Boudry, 
J-C.\,Brient, 
D.\,Jeans, 
P.\,Mora de Freitas, 
G.\,Musat, 
M.\,Reinhard, 
M.\,Ruan,  
H.\,Videau
\\ 
\begin{footnotesize}
\it Laboratoire Leprince-Ringuet (LLR)  -- \'{E}cole Polytechnique, CNRS/IN2P3, F-91128 Palaiseau, France
\end{footnotesize}

 M.\,Marcisovsky, 
 P.\,Sicho, 
 V.\,Vrba, 
 J.\,Zalesak 
\\ 
\begin{footnotesize}
\it
Institute of Physics, Academy of Sciences of the Czech Republic, Na Slovance 2,
CZ-18221 Prague 8, Czech Republic
\end{footnotesize}

B.\,Belhorma,
H.\,Ghazlane
\\ 
\begin{footnotesize}
\it
Centre National de l'Energie, des Sciences et des Techniques Nucl\'{e}aires, 
B.P. 1382, R.P. 10001, Rabat, Morocco
\end{footnotesize}

\end{center}
\renewcommand\notesname{\vspace*{-1.5cm}}
\theendnotes
\newpage

%% file: exposure.bbl
\begin{thebibliography}{99}
\bibitem{ild09} The ILD Concept Group.\\
{\em The International Large Detector - Letter of Intent.}\\
DESY 2009-87, Fermilab-Pub-09-682-E, KEK Report 2009-6, arXiv:1006.3396v1 [hep-ex]. 
\bibitem{sid-loi}
SiD Concept Group, H.~Aihara {\it et al.}.\\
{\em SiD Letter of Intent}.\\
arXiv:0911.0006 [physics.ins-det].

\bibitem{fourth}
A.~Mazzacane.\\
{\em The 4th Concept Detector for the ILC.}\\
NIM A {\bf 617} 173 (2010).\\
See the full text under: \url{http://www.4thconcept.org/4LoI.pdf.}

\bibitem{hre} A.~Holmes-Siedle and L.~Adams.\\
{\em Handbook of Radiation Effects.}\\
Oxford University Press, 1993, ISBN 0-19-856347-7.
\bibitem{imad} I.~Laktineh.\\
{\em CALICE Results and Future Plans}.\\
Presentation at the International Conference on High Energy Physics 2010, ICHEP2010. To appear in the proceedings.
\bibitem{calice1} CALICE Collaboration J.\,Repond \mbox{\it et al.}.\\
 {\em Design and electronics commissioning of the physics Prototype of a Si-W electromagnetic calorimeter for the International Linear Collider.}\\ 
 JINST {\bf 3} P08001 (2008), arXiv:0805.4833v2 [physics.ins-det]. 
\bibitem{dlt09}
C.~de la Taille.\\
{\em Front End Electronics in Calorimetry: From LHC to ILC}.\\
M\'emoire d'habilitation \`a encadrer des recherches, LAL-09-117.
\bibitem{calice2}
CALICE Collaboration,  J.~Repond \mbox{\it et al.}. \\
{\em Response of the CALICE Si-W electromagnetic calorimeter physics prototype to electrons}.\\
Nucl. Instr. and Methods A {\bf 608} 372 (2009), arXiv:0811.2354v1  [physics.ins-det].
\bibitem{calice3}
CALICE Collaboration,  C.~Adloff \mbox{\it et al.}. \\
{\em Study of the interactions of pions in the CALICE silicon-tungsten calorimeter prototype}.\\
JINST {\bf 5} P05007 (2010), arXiv:1004.4996 [physics.ins-det].
\bibitem{ATLAS-Note} R.~Zitoun.\\
{\em Study of Noise in the November 1998 barrel run.}\\
ATL-LARG-99-006.
\bibitem{ieee2000} R.~Everson and S.~Roberts.\\
{\em Inferring the Eigenvalues of Covariance Matrices from Limited, Noisy Data.}\\
IEEE Transactions on Signal Processing Vol.48, No. 7, 2083 (2000).
\bibitem{bro2007} R.~Bro, E.~Acar and T.G.~Kolda.\\
{\em Resolving the sign ambiguity in the singular value decomposition.}\\
Journal of Chemometrics, Volume {\bf 22}, Issue 3, 135 (2007). 
\bibitem{brandt}
S.~Brandt.\\
Datenanalyse.\\
BI-Wiss.-Verl., 1992, ISBN 3-411-03200.
\bibitem{cranmer} K.S.~Cranmer.\\
{\em Kernel Estimation in High-Energy Physics}.\\
Comput.Phys.Commun. {\bf 136} 198 (2001), arXiv:hep-ex/0011057v1.
\bibitem{bogdan} B.~Malaescu.\\
{\em Mesures pr\'ecises de sections efficaces $\epem\rightarrow{\rm Hadrons}$: tests du Mod\`ele Standard et applications en QCD}.\\
Th\`ese doctorale, LAL-10-113.
\bibitem{eudet} R.~P\"oschl.\\
{\em A large scale prototype for an SiW electromagnetic calorimeter for the ILC - EUDET module}.\\
Nucl. Instr. and Meth. A {\bf 617} 113 (2010).


\end{thebibliography}
